\documentclass[superscriptaddress,reprint,amsmath,amssymb,aps,floatfix]{revtex4-1}

\usepackage{amsmath,gensymb,textcomp,bm,dcolumn,eurosym,array,tabu,multirow,nicefrac,color,subfigure,graphicx,upgreek}
\usepackage[colorlinks, linkcolor=blue, citecolor=blue, urlcolor=blue, breaklinks=true]{hyperref}

\usepackage{mathpazo,times} 

\newcommand{\bra}[1]{\langle #1 \rvert}
\newcommand{\ket}[1]{\lvert #1 \rangle}

\usepackage[subsectionbib]{bibunits}
\usepackage{graphicx}
\usepackage{dcolumn}
\usepackage{bm,MnSymbol}
\usepackage{units}
\usepackage{psfrag}
\usepackage{float}
\usepackage{color}

\begin{document}
\title{Coherent generation of the complete high-dimensional Bell basis by adaptive pump modulation}

\author{Yuanyuan Chen}
\email{chenyy@xmu.edu.cn}
\affiliation{Department of Physics and Collaborative Innovation Center for Optoelectronic Semiconductors and Efficient Devices, Xiamen University, Xiamen 361005, China}
\author{\normalfont\textsuperscript{, \textcolor{blue}{\dag}} Wuhong Zhang}
\thanks{These authors contribute equally to this work.}
\affiliation{Department of Physics and Collaborative Innovation Center for Optoelectronic Semiconductors and Efficient Devices, Xiamen University, Xiamen 361005, China}
\author{Dongkai Zhang}
\affiliation{Department of Physics and Collaborative Innovation Center for Optoelectronic Semiconductors and Efficient Devices, Xiamen University, Xiamen 361005, China}
\author{Xiaodong Qiu}
\affiliation{Department of Physics and Collaborative Innovation Center for Optoelectronic Semiconductors and Efficient Devices, Xiamen University, Xiamen 361005, China}
\author{Lixiang Chen}
\email{chenlx@xmu.edu.cn}
\affiliation{Department of Physics and Collaborative Innovation Center for Optoelectronic Semiconductors and Efficient Devices, Xiamen University, Xiamen 361005, China}

\begin{abstract}
The Bell basis, a set of maximally entangled biphoton state, is a critical prerequisite towards quantum information processing, and many quantum applications have highlighted the requirement for the manipulation of high-dimensional Bell basis. While the Bell states can be created by using ingenious single-photon quantum gates, its implementation complexity in higher dimensions is significantly increased. Here we present an elaborate approach to show that the adaptive pump modulation enable the efficient preparation of Bell basis in arbitrary-dimensional Hilbert space. A complete set of four-dimensional orbital angular momentum Bell states are experimentally created, yielding high fidelities for certifying the entanglement dimensionality. Our strategy can be simply generalized to prepare more complex forms of quantum states even exploiting other physical degrees of freedom. Also, it can facilitate the use of high-dimensional entanglement in a variety of quantum protocols, in particular those requiring quantum dense coding.
\end{abstract}
\maketitle

\indent \emph{Introduction.}\rule[2pt]{8pt}{1pt} Quantum entanglement is an essential resource for a variety of quantum information applications \cite{horodecki2009quantum}, as well as for studying fundamental tests of quantum mechanics \cite{giustina2015significant}. Photon pairs entangled in high dimensions can offer more specific advantages over qubit systems and classical possibility, which usually facilitate the implementation of many complex quantum protocols \cite{mair2001entanglement,malik2016multi,dada2011experimental}. Since single photon can carry more information, high-dimensional entanglement is an enabling resource for dense coding, making them compelling for quantum communication with higher channel capacity \cite{luo2019quantum,barreiro2008beating}. The use of high-dimensionally entangled state can enhance the security of quantum communication by improving its robustness against noise and malicious eavesdropping attacks \cite{ecker2019overcoming,islam2017provably}. In addition, such states can potentially be used to enhance the processing capability and offer ever-increasing transmission capacity in quantum correlation imaging \cite{howland2013efficient}.

Engineering of a complete set of maximally entangled state, i.e., the so-called Bell basis, forms the foundation for many quantum information protocols. For instance, quantum teleporation \cite{wang2015quantum,luo2019quantum}, entanglement swapping \cite{zhang2017simultaneous,takeda2015entanglement} and all-photonic quantum repeater \cite{hasegawa2019experimental} build on the deterministic discrimination of Bell states. In quantum communication via dense coding \cite{barreiro2008beating}, the generation, manipulation and measurement of Bell states is an absolute necessity since they act as the target for encoding information. Besides, with the recent availability of quantum computation, the technology of manipulating Bell states in a quantum computer has been brought into focus, which in turn pushes forward the development of quantum computation \cite{arute2019quantum}. While two-dimensional Bell states entangled in polarization, time-bin or orbital angular momentum (OAM) have already been widely used in quantum applications, the versatile engineering of high-dimensional Bell basis is still a formidable challenge \cite{zhang2016engineering,kong2019manipulation}.

To tackle this issue, Wang \textit{et al.} presented a well-designed scheme for experimental generation of a complete basis of two-photon four-dimensional maximally entangled quantum states \cite{wang2017generation}. By performing generalized Pauli-X and Pauli-Z transformation on an initial entangled state, 16 Bell states in the OAM degree of freedom were generated. The high-dimensional X-gate, corresponding to a cyclic transformation, was required to modulate the relation between different spatial modes \cite{babazadeh2017high}. Since the implementation complexity of cyclic transformation in high dimensions is exponentially increased, the creation of arbitrary-dimensional Bell basis puts a brake on the development of high-dimensional quantum information processing \cite{gao2019arbitrary}.

Recent years have also witnessed some clever combinations of structured light with quantum state engineering. In the generation of quantum entanglement, the use of pump with a spatially structured modulation in spontaneous parametric down conversion (SPDC) process enables a new form of quantum pattern recognition \cite{qiu2019structured}, or offers a new method of engineering the maximally entangled photon pairs \cite{chen2018polarization,chen2020verification,kovlakov2018quantum,kovlakov2017spatial,liu2018coherent}. However, the preparation of the complete arbitrary-dimensional Bell basis still remains relatively unexplored. Here we further explore the feature of spatial modulation of pump beam for presenting a concise yet efficient method to prepare the complete high-dimensional Bell basis. In our scheme, by harnessing the input pump beam skillfully, we are able to modulate the on-demand relation between different spatial modes of the down-converted photons, and thus significantly demolishing the barriers of cyclic transformation. A high-dimensional Pauli-Z gate is applied to one of the entangled photons to fulfill the task of mode-dependent phase transformation, corresponding to modulate the relative phase between superposition states.

We experimentally create 16 Bell states that form a complete basis of four-dimensional Hilbert space. To quantify the quality of our generated quantum states, we perform quantum state tomography to reconstruct their density matrices that allow for measuring their overlap to the ideal Bell basis. From the calculated fidelity, we also certify the entanglement dimensionality for paving the way to its application in quantum information processing. Backed by our proof-of-concept experiment, it suggests that this approach can provide great advantages on implementation complexity and entanglement quality over high-dimensional single-photon quantum gates.

\indent \emph{Theoretical scheme.}\rule[2pt]{8pt}{1pt} The general formulation of $d$-dimensional Bell states of a bipartite system can be expressed as
\begin{equation}
\label{eq:general bell basis}
\ket{\varphi}_{m,n}=\frac{1}{\sqrt{d}}\sum_{k=0}^{d-1}e^{\frac{2\pi}{d}nk}\ket{k}_A\ket{m\oplus k}_B,
\end{equation}
where $m\oplus k=(m+k)$ mod $d$, the variable $m$ defines the correlation between pairs of photons, while the variable $n$ defines the relative phase relationship between the superposition states \cite{bennett1993teleporting}. As a direct result of mutual orthogonality between two arbitrary Bell states, we note that Bell states $\ket{\varphi_{m,0}}$ can be prepared by performing Pauli-X gate on an initial entangled state $\ket{\varphi_{0,0}}$. The generality of this statement is remarkable: two superposition states with $\pm 1$ OAM quanta cannot exist in one Bell state, i.e., superposition states $\ket{j}_A\ket{k}_B$ and $\ket{j\pm1}_A\ket{k}_B$ or $\ket{j}_A\ket{k\pm1}_B$ cannot simultaneously appear in any $\ket{\varphi_{m,n}}$. To tackle the issue of generating arbitrary Bell states with the assistance of pump modulation, we reconstruct a more elaborate set of the complete Bell basis by following the preceding rule. Taking the mode-dependent phase modulation into consideration, we can reform the quantum states as
\begin{equation}
\label{eq:new bell basis}
\ket{\psi}_{m,n}=\frac{1}{\sqrt{d}}\sum_{k=0}^{d-1}e^{\frac{2\pi}{d}nk}\ket{k}_A\ket{m\ominus k}_B,
\end{equation}
where $m\ominus k=(m-k)$ mod $d$.

Since the superposition state $\ket{k_A}\ket{k_B}$ of down-converted photons demonstrated in Eq. \eqref{eq:new bell basis} satisfy $k_A+k_B=m$ mod $d$, it is clear that the flexible control of correlation can be readily implemented by manipulating the input pump beam adaptively. Accordingly, the appropriate choice of pump state is of the utmost importance. In addition, the conservation of energy and momentum is an inherent nature of SPDC process, our method can be applied in a variety of physical degrees of freedom such as polarization, orbital angular momentum, path and frequency.

\begin{figure}[!t]
\centering
\includegraphics[width=\linewidth]{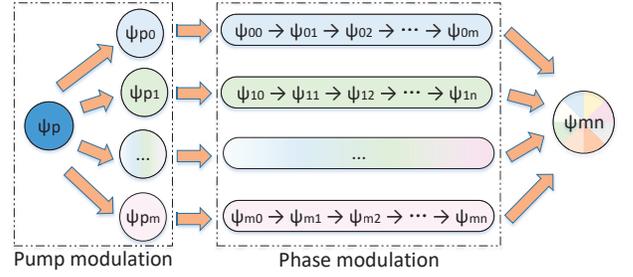}
\caption{By employing a modulation on the input pump beam and a mode-dependent phase transformation on one photon of the down-converted pairs, a complete set of maximally entangled high-dimensional Bell states can be created.}
\label{figure_1}
\end{figure}

Here we focus on the generation of the complete four-dimensional Bell basis in OAM degree of freedom, i.e., $d=m=n=4$. For the preparation of these Bell states, we are required to conduct two basic operations: adaptive pump modulation ($\ket{\psi}_p\rightarrow\ket{\psi}_{p0},\ket{\psi}_{p1},\ket{\psi}_{p2},\ket{\psi}_{p3}$) and mode-dependent phase transformation ($\ket{\psi}_{m,0}\rightarrow\ket{\psi}_{m,1},\ket{\psi}_{m,2},\ket{\psi}_{m,3}$) as shown in Fig. \ref{figure_1}.
By engineering the input pump beam of SPDC process \cite{walborn2004entanglement}, the spatial spectrum of down-converted photons can be purposefully manipulated. If the pump is in a superposition of multiple OAM states $\sum_{L_p} C_L\ket{L_p}$ ($C_L$ is a complex number), photons from this strong pump laser spontaneously decay into signal and idler photons, with conservation of momentum resulting in OAM entanglement as
\begin{equation}
\label{eq: spdc}
\ket{\phi}=\sum_{L_p}C_L\sum_{\ell=-\infty}^{\infty}c_\ell\ket{\ell}_s\ket{L_p-\ell}_i,
\end{equation}
where $c_\ell$ represents a coincidence probability for finding one signal photon in OAM state $\ket{\ell}_s$ and the corresponding idler photon in OAM state $\ket{L_p-\ell}_i$. Since the inherent unbalancing of the created modes can be overcome with modulation of pump beam and Procrustean filtering technique \cite{vaziri2003concentration}, the resultant quantum states can be written as
\begin{equation}
\begin{split}
C_{-2}\ket{-2}_{p}+C_2\ket{2}_{p}\rightarrow&(\ket{-1,-1}+\ket{0,2}+\ket{1,1}+\ket{2,0})/2\\
C_{-1}\ket{-1}_{p}+C_3\ket{3}_{p}\rightarrow&(\ket{-1,0}+\ket{0,-1}+\ket{1,2}+\ket{2,1})/2\\
C_0\ket{0}_{p}+C_4\ket{4}_{p}\rightarrow&(\ket{-1,1}+\ket{0,0}+\ket{1,-1}+\ket{2,2})/2\\
\ket{1}_{p}\rightarrow&(\ket{-1,2}+\ket{0,1}+\ket{1,0}+\ket{2,-1})/2,
\end{split}
\end{equation}
where $C_{L}$ represent the probability amplitudes of superposition state that fulfilling the normalization condition. Thus, the initial entangled states of each group $\ket{\psi}_{m,0}$ are created without any requirement of high-dimensional single-photon quantum gate.

The phase modulation can be applied with a Dove prism (DP) in signal or idler path. Generally, a DP oriented at an angle $\alpha$ would introduce a relative phase as $\ket{\ell}\rightarrow exp(i2\ell\alpha)\ket{\ell}$ that determined by the OAM value $\ell$ of the incident photon and the rotation angle $\alpha$ of the prism \cite{gonzalez2006dove,abouraddy2012implementing}. The effect of this operation on the state can be expressed as
\begin{equation}
\ket{\phi}\rightarrow\ket{\psi}=\frac{1}{\sqrt{d}}\sum_{\ell=0}^{d-1}exp(i2\ell\alpha)\ket{\ell}_A\ket{\ell}_B.
\end{equation}
By orienting the DP at different angles ($\alpha=0$, $\pi/4$, $\pi/2$, $3\pi/4$), the corresponding bell states in each group are readily generated as a result of mode-dependent phase transformation of $\ket{\psi}_{m,0}\xrightarrow{DP(\alpha)}\ket{\psi}_{m,1/2/3}$. Together, the engineering of pump beam and relative phase on signal or idler photon enables us to create 16 Bell states that forming a complete Bell basis in four-dimensional Hilbert space.

To quantify the entanglement dimensionality of the produced state, we measure two-photon correlations in a set of mutually unbiased bases, and the measurement results allow us to obtain the parameters in a density matrix by using quantum state tomography \cite{thew2002qudit,agnew2011tomography}. Then the fidelity is estimated by calculating the overlap of the experimentally measured state to an ideal Bell basis, which is a most commonly used evaluator for certifying the dimensions of entanglement \cite{friis2019entanglement}.

\indent \emph{Experimental implementation.}\rule[2pt]{8pt}{1pt}
\begin{figure}[!t]
\centering
\includegraphics[width=0.9\linewidth]{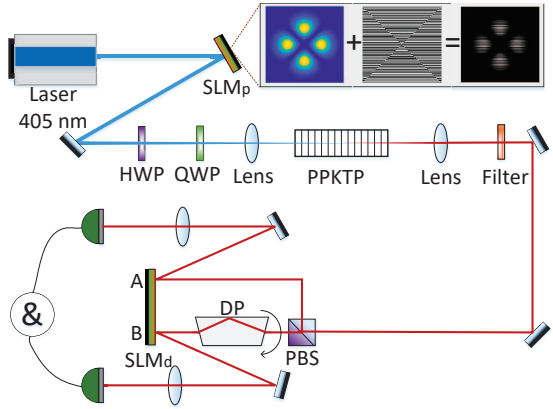}
\caption{Experimental setup for generating and detecting high-dimensional Bell states. SLM: spatial light modulator; HWP: half-wave plate; QWP: quarter-wave plate; ppKTP: type-II periodically poled potassium titanyl phosphate crystal; PBS: polarizing beam splitter; DP: Dove prism. SLM$_\text{p}$ and SLM$_\text{d}$ are spatial ultraviolet and near-infrared light modulators used to modulate the pump beam and measure the OAM-entangled state. }
\label{figure_2}
\end{figure}
As shown in Fig. \ref{figure_2}, we generate high-dimensional OAM entanglement via spontaneous parametric down conversion in a 5-mm type-II ppKTP crystal pumped with a continuous wave laser at a wavelength of $\unit[405]{nm}$. The desired input pump in a specific superposition state can be prepared by using a ultraviolet spatial light modulator (SLM), which allows us to tune the pump parameters in a more flexible way. Then a long-pass filter is used to eliminate the remaining pump light and noise. After being guided to a PBS, the entangled biphotons are routed into distinct spatial modes, so-called signal and idler photons. A rotatable DP is placed in the signal path for fulfilling the task of phase modulation. The projective measurements are performed by using a near-infrared SLM and single mode fibers. Finally, the down-converted photons are detected by silicon avalanche photon diodes, and twofold events are identified using a fast electronic AND gate when two photons arrive at the detectors within a coincidence window of $\sim\unit[1]{ns}$.
\begin{figure}[!t]
\centering
\subfigure[]{
\label{Fig3.sub.1}
\includegraphics[width=0.48\linewidth]{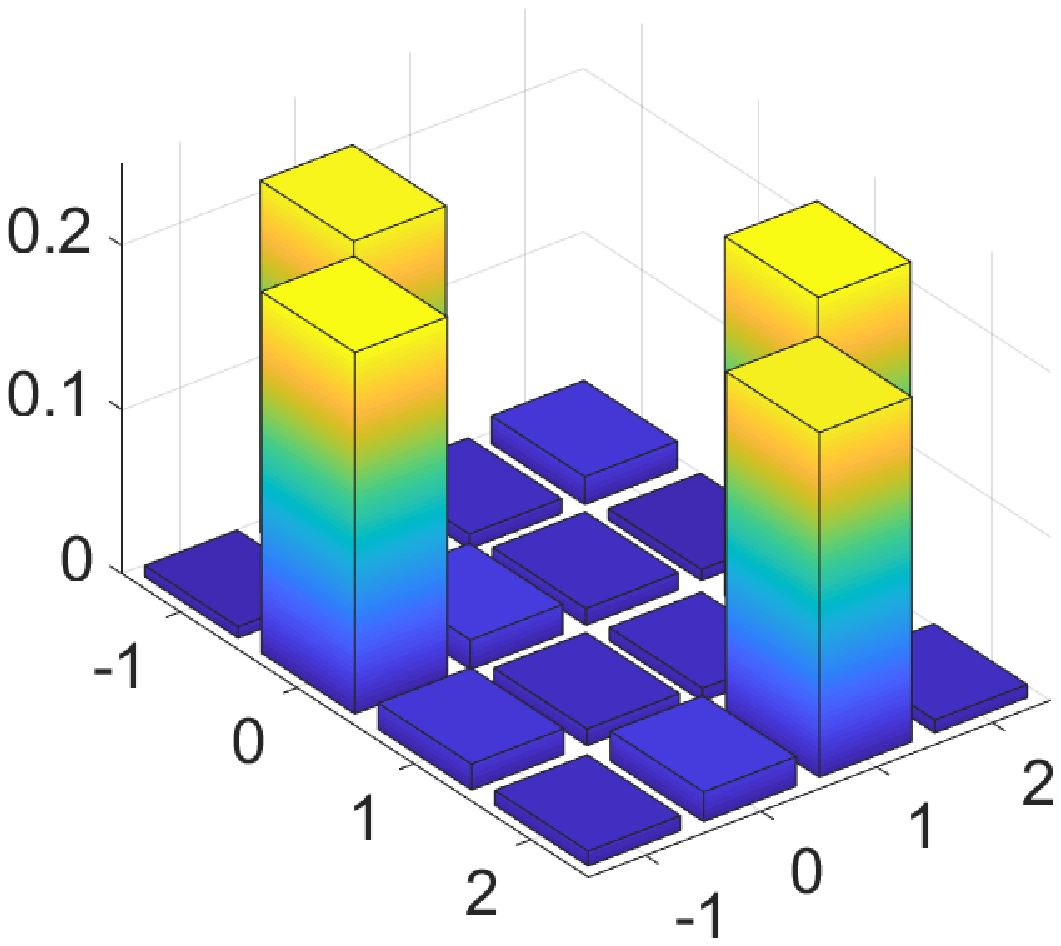}}
\subfigure[]{
\label{Fig3.sub.2}
\includegraphics[width=0.48\linewidth]{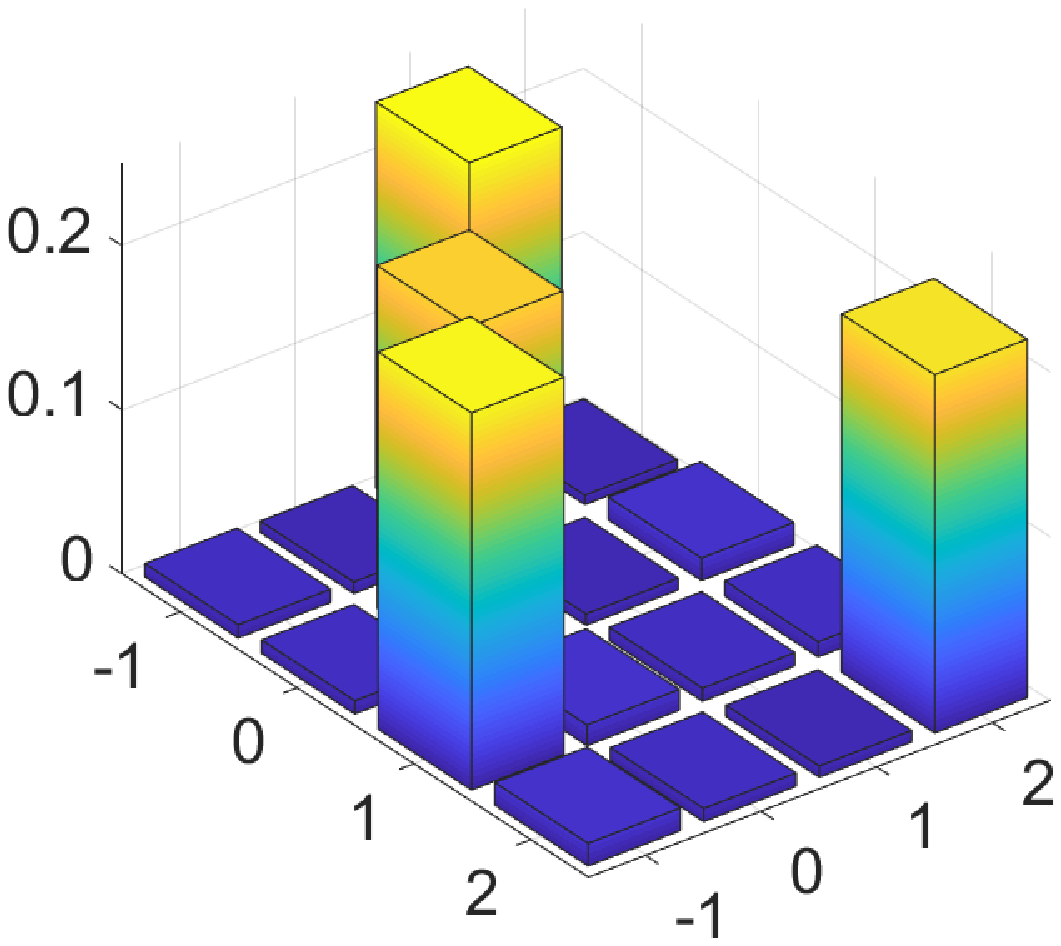}}
\subfigure[]{
\label{Fig3.sub.3}
\includegraphics[width=0.48\linewidth]{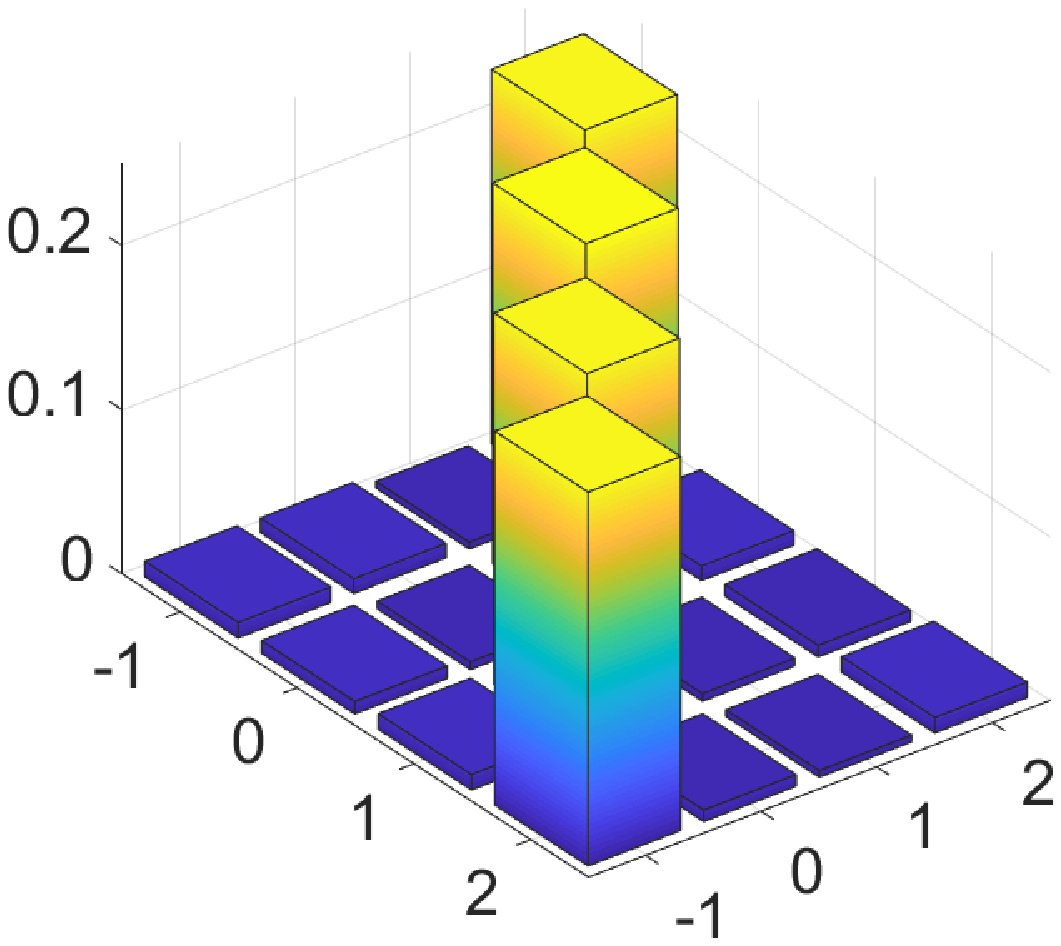}}
\subfigure[]{
\label{Fig3.sub.4}
\includegraphics[width=0.48\linewidth]{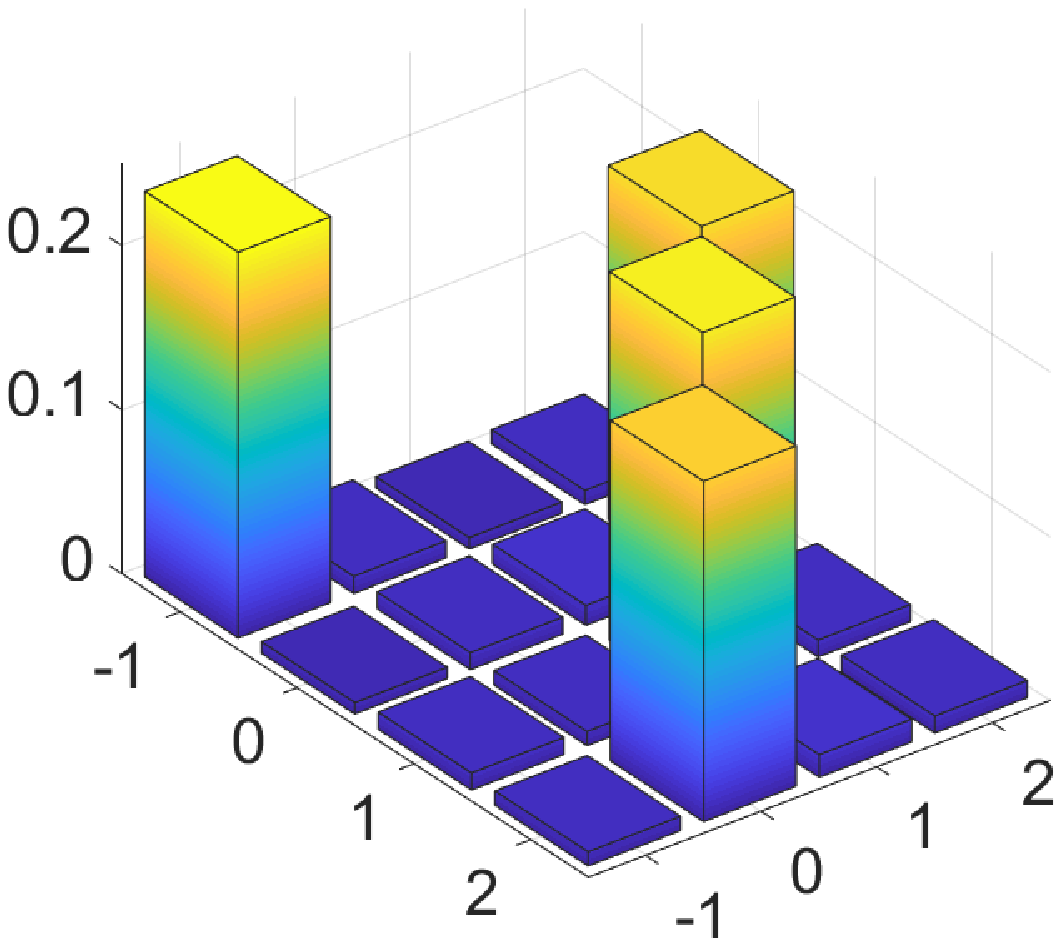}}
\caption{The spatial spectrum distribution for initial entangled state (a) $\ket{\psi}_{0,0}$, (b) $\ket{\psi}_{1,0}$, (c) $\ket{\psi}_{2,0}$ and (d) $\ket{\psi}_{3,0}$.}
\label{figure_3}
\end{figure}

Through the technique of modulating the input pump beam adaptively, the initial spatially entangled state $\ket{\psi}_{0/1/2/3,0}$ is readily generated (see Fig. \ref{figure_3}). By applying the mode-dependent phase transformation, the remaining quantum state $\ket{\psi}_{m,0/1/2/3}$ can be created. According to quantum state tomography, we are able to construct the density matrix $\rho_{exp}$ by minimizing the Chi-square quantity \cite{jack2009precise}, which can be efficiently solved via semidefinite programming (SDP), a class of convex optimization problems. All the reconstructed density matrices of experimentally generated quantum states are outlined in Appendix \textcolor{blue}{E}.

\begin{figure}[!t]
\centering
\includegraphics[width=\linewidth]{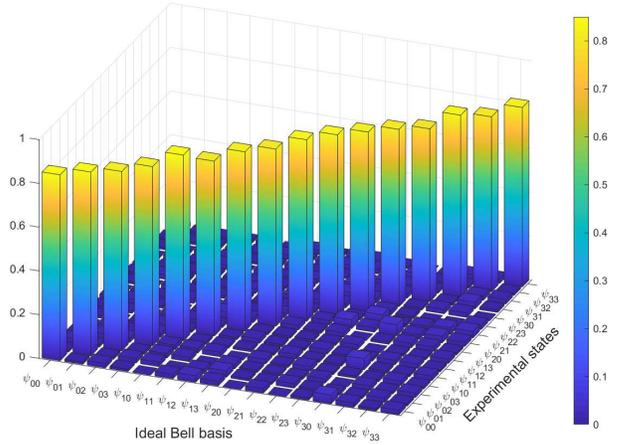}
\caption{Fidelity of experimentally generated state to the ideal Bell basis. The diagonal elements indicate the overlap between the experimentally generated states and its corresponding target states.}
\label{figure_4}
\end{figure}

To quantify the quality of these generated Bell state, both of a fidelity to the theoretical expected Bell basis and a witness of four-dimensional entanglement are analyzed. The fidelity to the ideal Bell basis $\ket{\psi}_{m,n}$ follows from $F_{exp}=Tr(\rho_{exp}\ket{\psi}_{m,n}\bra{\psi}_{m,n})$, and it reveals the overlap between the experimental state and the ideal Bell basis. Figure \ref{figure_4} shows the fidelity of experimentally created states to each of the complete four-dimensional Bell basis.
The average fidelity to the ideal Bell basis for the 16 experimentally generated state is $0.821\pm0.0223$. The decreased measured fidelity can be attributed to the intermodal crosstalk between different spatial modes, imperfect misalignment and nonideal optical components.

\begin{figure}[!t]
\centering
\includegraphics[width=\linewidth]{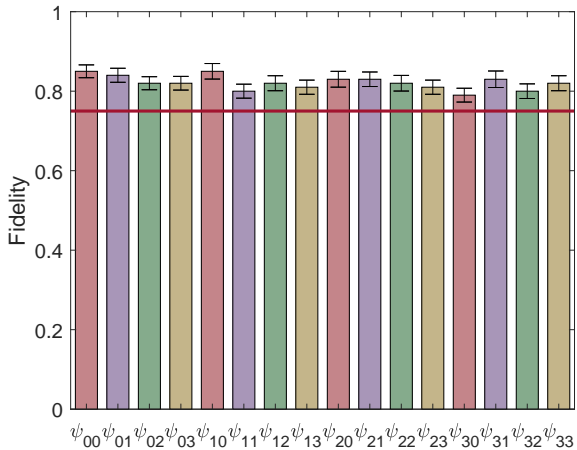}
\caption{Fidelity witness. All of the generated 16 Bell states in our experiment exceed the theoretical upper bound for the overlap with a three-dimensionally entangled state (red line).}
\label{figure_5}
\end{figure}
The witness enables us to certify the entanglement dimensionality of our produced states (see Appendix \textcolor{blue}{D}). By using bipartite entanglement witness for four-dimensional systems, we note that the theoretical maximal overlap of an arbitrary three-dimensional quantum state with a maximally entangled four-dimensional state is $\bar{F}_{max}=75\%$. It means that if our generated state in the experiment exceeds this upper bound, the state is certified to be four-dimensionally entangled. Figure \ref{figure_5} shows the measured fidelity witnesses for all 16 Bell states, each of which individually exceeds this upper bound. Thus we are sufficient to certify that all of these Bell state generated in our experiment are at least four-dimensionally entangled.

\indent \emph{Discussion.}\rule[2pt]{8pt}{1pt} To facilitate the high-dimensional quantum dense coding, our approach can generate a complete set of arbitrary-dimensional Bell states in an on-demand manner. In theory, we can encode 4 bits of information as a direct result of the utilization of two-photon four-dimensional Bell basis. The method can be extended to higher dimensional quantum systems such that higher information capacity is obtained by solely using single photon encoding. But in experiment, the measured results in this work allows us to obtain a mutual information of 2.8 bits. This value exceeds the theoretical upper limit for two-photon two-dimensionally entangled state of 2 bits. In the applications of high-dimensional encoding, the unambiguously discrimination of all Bell states is essential. However, the theory has proven that it is impossible to distinguish all Bell states with just linear optics. To tackle this challenge, several solutions have been proposed to implement the complete Bell-state measurement in high dimensions, such as nonlinear optics \cite{kim2001quantum} and auxiliary degrees of freedom \cite{luo2019quantum}. We believe high-dimensional entanglement can be an ideal candidate for quantum communication in the foreseeable future.

While our experimental realization is based on photons' spatial mode, the overall scheme could also be applied to other degrees of freedom, e.g., frequency or path. Another promising line of inquiry could address the using of this approach in engineering more complex forms of quantum states like high-dimensional Dicke state and multi-qubit cluster state. We also envisage the use of this approach in manipulating multiphoton experiment, where the generation of multiphoton entanglement with high fidelity is still an open challenge.

In conclusion, we hope our results could inspire new experimental configuration for the generation, manipulation and detection of complex forms of high-dimensional entanglement. We believe this flexible method for preparing arbitrary entangled quantum state paves the way to next generation of quantum information processing applications.

\section*{Acknowledgements}
This work is supported by the National Natural Science Foundation of China (91636109, 11904303), the Fundamental Research Funds for the Central Universities at Xiamen University (20720190054, 20720190057), the Natural Science Foundation of Fujian Province of China for Distinguished Young Scientists (2015J06002), and the program for New Century Excellent Talents in University of China (NCET-13-0495).

\bibliography{apssamp}

\clearpage
\newpage
\section*{Appendix}
\subsection{Deviations from ideal Bell basis}
Here we mainly ascribe the deviations of the experimentally created quantum states from the ideal Bell basis to the non-equal distribution of spatial modes and crosstalk between adjacent modes. The non-flat OAM distribution of down-converted photon pairs can be attributed to two main effects: imperfect superposition state of the input pump and inherent characteristic of SPDC process. As a result, the initially generated state is not maximally entangled. In order to obtain high-dimensional entanglement with high quality, we harness the parameter $C_L$ in Eq. \textcolor{blue}{(3)} skillfully such that the unbalancing caused by the input pump is eliminated. In addition, Procrustean filter technique is utilized to eliminate the inherent unbalancing of SPDC process at the cost of efficiency. 
\subsection{Quantum state tomography for four-dimensional entanglement}
Quantum state tomography is the process to reconstruct a density matrix of an unknown quantum state by performing a set of measurements on identical copies of the relevant quantum state. Here,we use two groups of quantum states as the measurement basis: the pure OAM states,
\begin{equation}
\ket{\Psi}_\ell=\ket{\ell},
\end{equation}
and superpositions of just two of these pure states,
\begin{equation}
\ket{\Psi}_{\alpha,\ell_1,\ell_2}=\frac{1}{\sqrt{2}}(\ket{\ell_1}+e^{i\alpha}\ket{\ell_2}),
\end{equation}
where $\alpha=0$, $\pi/2$, $\pi$, $3\pi/2$ and $\ell_1<\ell_2$. In this work, we create two-photon four-dimensional entanglement, resulting in a $16\times16$ density matrix. In order to determine its parameters, we are required to minimize the deviation from the ideal Bell basis, which can be revealed by the Chi-square quantity:
\begin{equation}
\begin{split}
\chi^2=\sum_{i=1}^{N^2}\frac{(p_i^e-p_i^t)^2}{p_i^t},
\end{split}
\end{equation}
where $p_i^t$ are theoretically probabilities calculated from the ideal density matrix, and $p_i^e$ are the measured probabilities in the experiment. Due to the requirement of describing a real physical system, the predicted density matrix must subject to that its eigenvalues are nonnegative and its trace equals to unity. Thus, SDP method can enable us to obtain an optimized density matrix that is extracted from precise knowledge about the experimental measurement results.

\subsection{Overlap between states}
Table \ref{tab1} presents the data from which Fig. \textcolor{blue}{3} is created, which reveals the overlap between experimentally generated quantum state and the theoretically expected Bell basis.
\begin{table*}[hbt]
\begin{tabular}{p{0.9cm}<{\centering}p{0.9cm}<{\centering}p{0.9cm}<{\centering}p{0.9cm}<{\centering}p{0.9cm}<{\centering}p{0.9cm}<{\centering}p{0.9cm}<{\centering}p{0.9cm}<{\centering}p{0.9cm}<{\centering}p{0.9cm}<{\centering}p{0.9cm}<{\centering}p{0.9cm}<{\centering}p{0.9cm}<{\centering}p{0.9cm}<{\centering}p{0.9cm}<{\centering}p{0.9cm}<{\centering}p{0.9cm}<{\centering}p{0.9cm}<{\centering}}
\hline
\hline
& $\ket{\Psi}_{0,0}$ & $\ket{\Psi}_{1,0}$ & $\ket{\Psi}_{2,0}$ & $\ket{\Psi}_{3,0}$ & $\ket{\Psi}_{0,1}$ & $\ket{\Psi}_{1,1}$ & $\ket{\Psi}_{2,1}$ & $\ket{\Psi}_{3,1}$ & $\ket{\Psi}_{0,2}$ & $\ket{\Psi}_{1,2}$ & $\ket{\Psi}_{2,2}$ & $\ket{\Psi}_{3,2}$ & $\ket{\Psi}_{0,3}$ & $\ket{\Psi}_{1,3}$ & $\ket{\Psi}_{2,3}$ & $\ket{\Psi}_{3,3}$\\
\hline
$\ket{\psi}_{0,0}$ & 0.85 & 0.01 & 0.02 & 0.01 & 0.00 & 0.02 & 0.03 & 0.02 & 0.00 & 0.02 & 0.00 & 0.01 & 0.04 & 0.02 & 0.01 & 0.02\\
$\ket{\psi}_{1,0}$ & 0.02 & 0.84 & 0.01 & 0.02 & 0.02 & 0.00 & 0.02 & 0.03 & 0.01 & 0.01 & 0.02 & 0.01 & 0.01 & 0.01 & 0.02 & 0.01\\
$\ket{\psi}_{2,0}$ & 0.02 & 0.01 & 0.82 & 0.01 & 0.02 & 0.02 & 0.00 & 0.02 & 0.01 & 0.01 & 0.01 & 0.02 & 0.00 & 0.02 & 0.01 & 0.02\\
$\ket{\psi}_{3,0}$ & 0.01 & 0.02 & 0.02 & 0.82 & 0.02 & 0.02 & 0.02 & 0.00 & 0.02 & 0.01 & 0.02 & 0.01 & 0.01 & 0.01 & 0.02 & 0.01\\
$\ket{\psi}_{0,1}$ & 0.00 & 0.02 & 0.01 & 0.02 & 0.85 & 0.01 & 0.01 & 0.01 & 0.01 & 0.02 & 0.01 & 0.01 & 0.05 & 0.01 & 0.01 & 0.02\\
$\ket{\psi}_{1,1}$ & 0.01 & 0.00 & 0.01 & 0.02 & 0.01 & 0.80 & 0.01 & 0.02 & 0.01 & 0.01 & 0.02 & 0.01 & 0.01 & 0.02 & 0.02 & 0.01\\
$\ket{\psi}_{2,1}$ & 0.02 & 0.01 & 0.00 & 0.02 & 0.02 & 0.01 & 0.82 & 0.01 & 0.01 & 0.01 & 0.01 & 0.02 & 0.00 & 0.01 & 0.01 & 0.02\\
$\ket{\psi}_{3,1}$ & 0.01 & 0.02 & 0.01 & 0.00 & 0.01 & 0.03 & 0.02 & 0.81 & 0.02 & 0.01 & 0.02 & 0.00 & 0.01 & 0.02 & 0.02 & 0.01\\
$\ket{\psi}_{0,2}$ & 0.00 & 0.02 & 0.01 & 0.02 & 0.00 & 0.02 & 0.01 & 0.02 & 0.83 & 0.02 & 0.04 & 0.02 & 0.05 & 0.02 & 0.03 & 0.02\\
$\ket{\psi}_{1,2}$ & 0.01 & 0.00 & 0.02 & 0.01 & 0.01 & 0.01 & 0.01 & 0.01 & 0.01 & 0.83 & 0.01 & 0.04 & 0.01 & 0.02 & 0.03 & 0.03\\
$\ket{\psi}_{2,2}$ & 0.01 & 0.02 & 0.01 & 0.01 & 0.01 & 0.02 & 0.01 & 0.02 & 0.02 & 0.02 & 0.82 & 0.02 & 0.01 & 0.00 & 0.00 & 0.02\\
$\ket{\psi}_{3,2}$ & 0.01 & 0.01 & 0.01 & 0.01 & 0.01 & 0.01 & 0.01 & 0.01 & 0.01 & 0.03 & 0.02 & 0.81 & 0.01 & 0.02 & 0.02 & 0.00\\
$\ket{\psi}_{0,3}$ & 0.00 & 0.02 & 0.01 & 0.02 & 0.00 & 0.02 & 0.01 & 0.02 & 0.00 & 0.03 & 0.01 & 0.02 & 0.79 & 0.01 & 0.03 & 0.01\\
$\ket{\psi}_{1,3}$ & 0.01 & 0.00 & 0.02 & 0.01 & 0.01 & 0.01 & 0.02 & 0.01 & 0.01 & 0.00 & 0.02 & 0.02 & 0.01 & 0.83 & 0.02 & 0.02\\
$\ket{\psi}_{2,3}$ & 0.01 & 0.01 & 0.01 & 0.01 & 0.01 & 0.02 & 0.01 & 0.02 & 0.02 & 0.02 & 0.00 & 0.03 & 0.01 & 0.01 & 0.80 & 0.01\\
$\ket{\psi}_{3,3}$ & 0.01 & 0.01 & 0.01 & 0.01 & 0.02 & 0.00 & 0.02 & 0.01 & 0.03 & 0.02 & 0.01 & 0.00 & 0.00 & 0.02 & 0.02 & 0.82\\
\hline
\end{tabular}
\caption{Overlap between experimentally measured states and theoretically expected states.}
\label{tab1}
\end{table*}
\subsection{Certification of high-dimensional entanglement}
Entanglement dimensionality represents a minimum number of levels that is required to faithfully describe a quantum state. For a $d\times d$-dimensional quantum state with a fidelity of $F$ to the expected entangled state, a lower bound of its entanglement dimensionality is expressed as
\begin{equation}
d_\text{ent} = \max\left\{k\,;\,F\leq\frac{k-1}{d}\right\},
\end{equation}
which has been proved. As a direct result, we can estimate the dimensionality of arbitrary entanglement from its fidelity. By using the above inequality, all of the measured fidelities in the experiment exceed the bound for a four-dimensional entangled state, which certify $d_\text{ent} = 4$. It also means that our generated quantum state can only be fully described with a $4\times4$-dimensionally entangled state.
\subsection{Density matrices of experimentally created entanglement states}
Figure \ref{figure_6} outlines the real and imaginary parts of the estimated density matrices for 16 experimentally measured Bell states.
\begin{figure*}[!t]
\centering
\subfigure[]{
\label{Fig6.sub.1}
\includegraphics[width=0.24\linewidth]{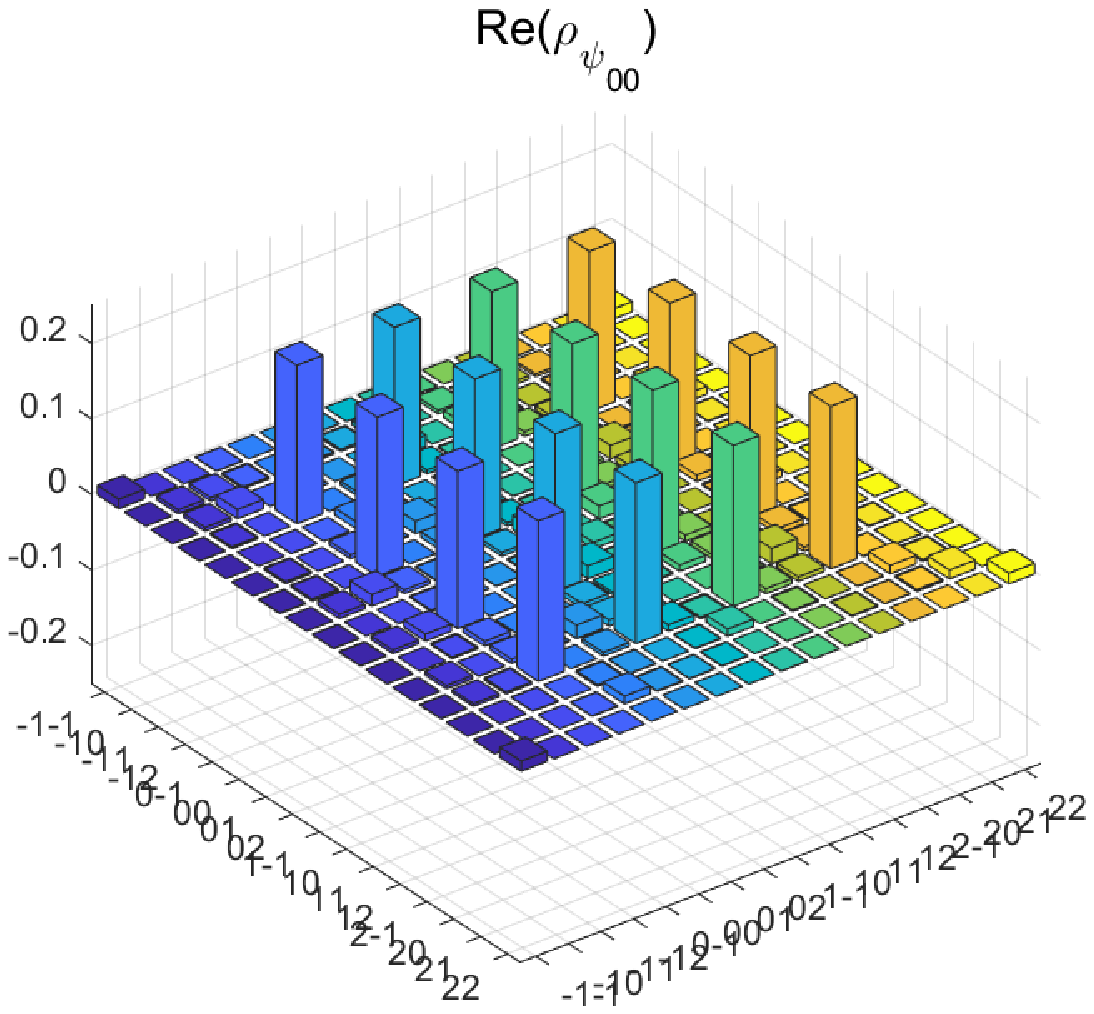}}
\subfigure[]{
\label{Fig6.sub.2}
\includegraphics[width=0.24\linewidth]{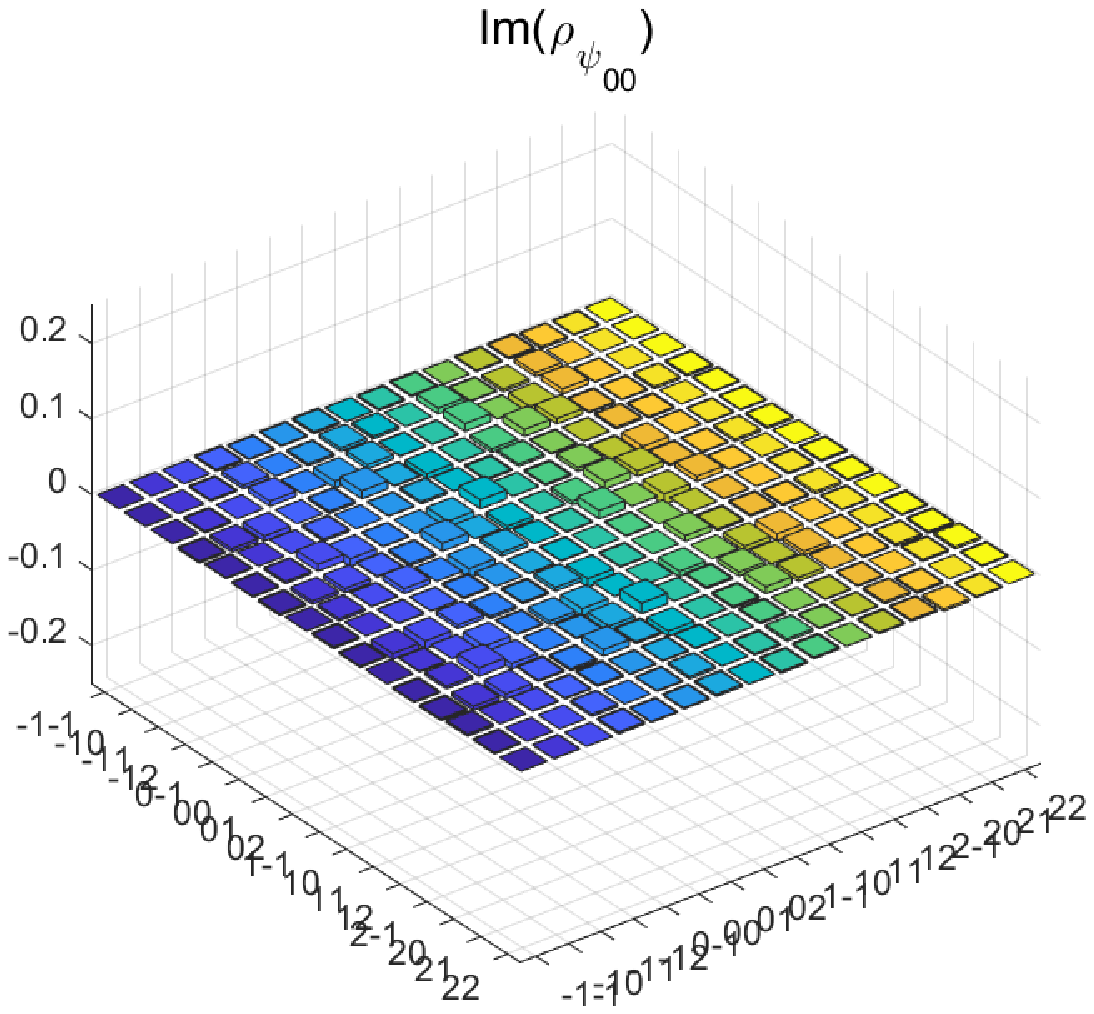}}
\subfigure[]{
\label{Fig6.sub.3}
\includegraphics[width=0.24\linewidth]{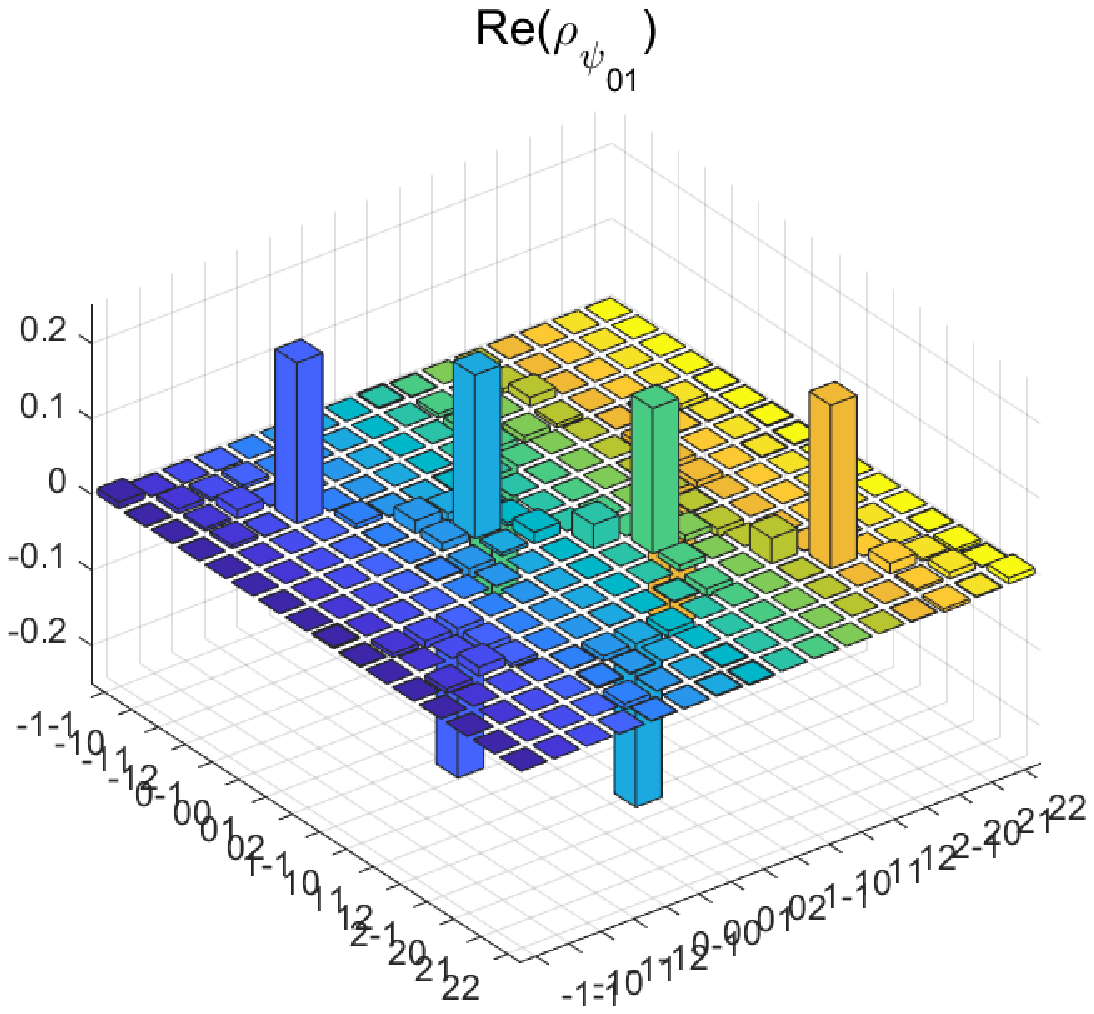}}
\subfigure[]{
\label{Fig6.sub.4}
\includegraphics[width=0.24\linewidth]{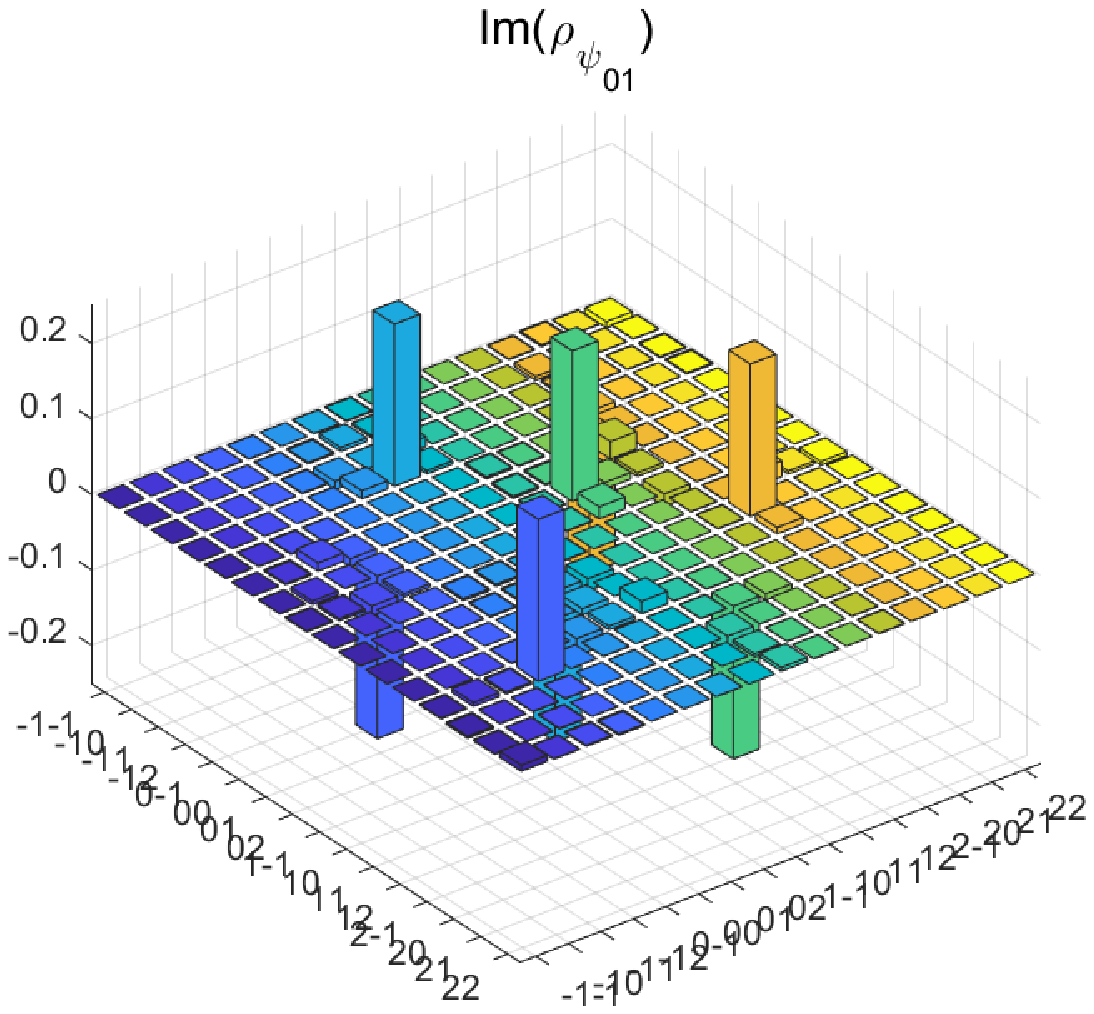}}
\subfigure[]{
\label{Fig6.sub.5}
\includegraphics[width=0.24\linewidth]{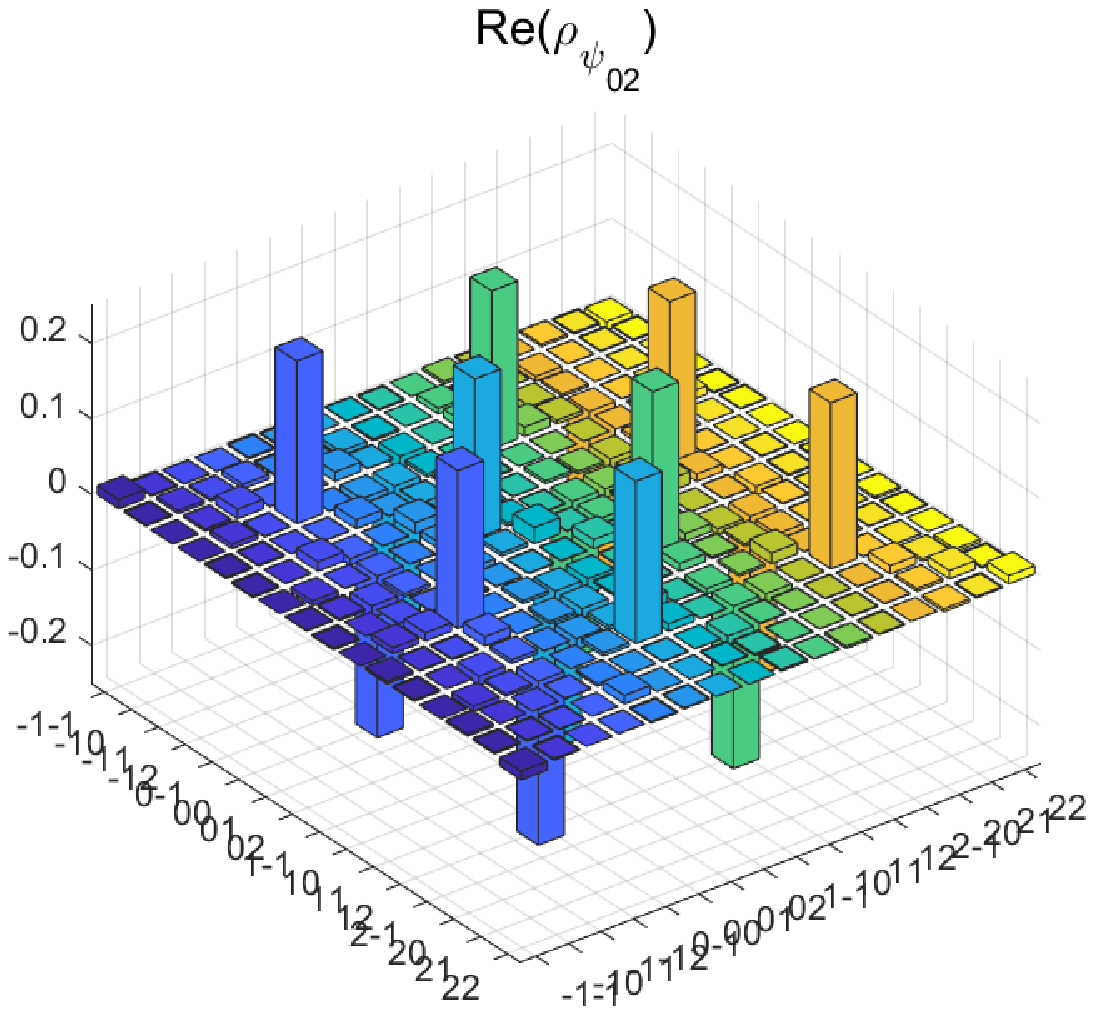}}
\subfigure[]{
\label{Fig6.sub.6}
\includegraphics[width=0.24\linewidth]{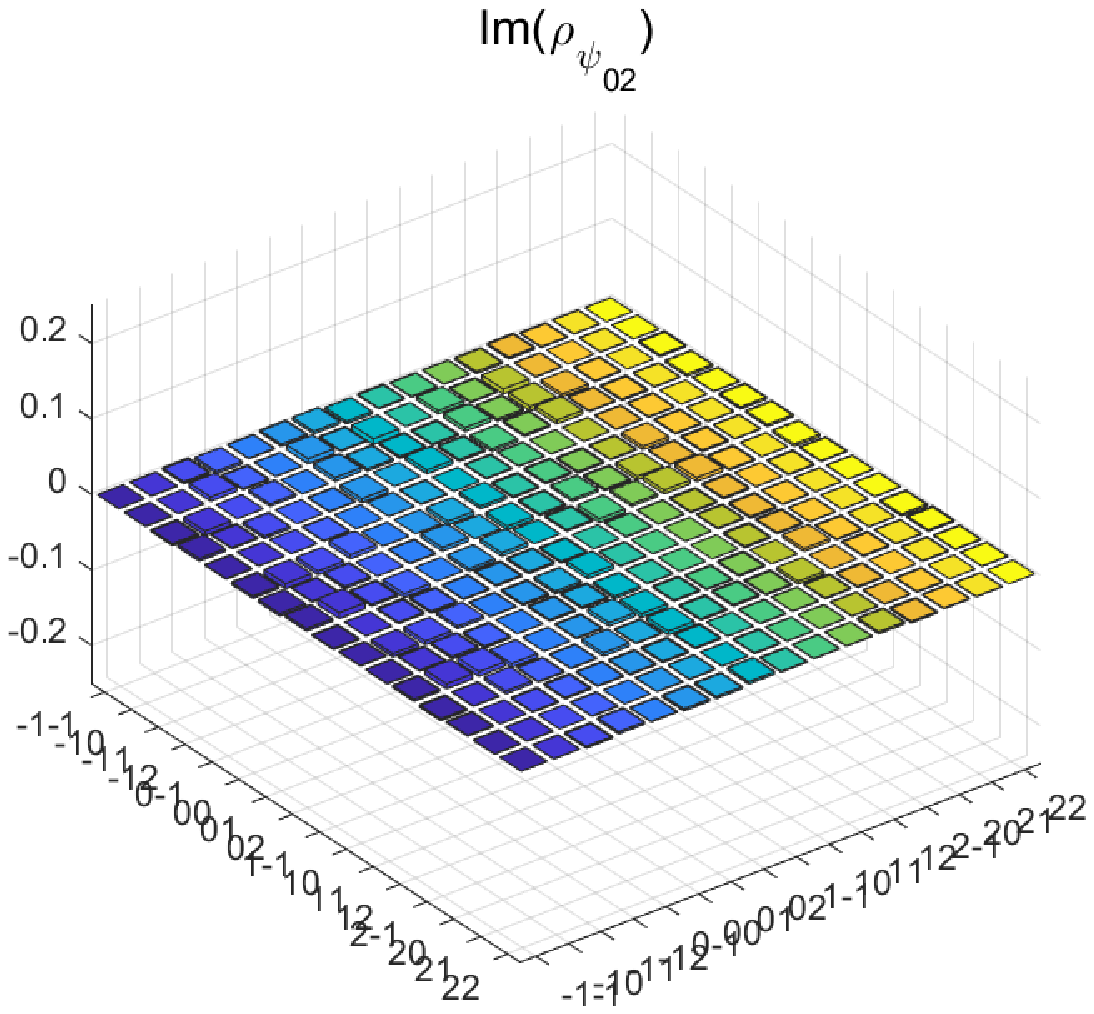}}
\subfigure[]{
\label{Fig6.sub.7}
\includegraphics[width=0.24\linewidth]{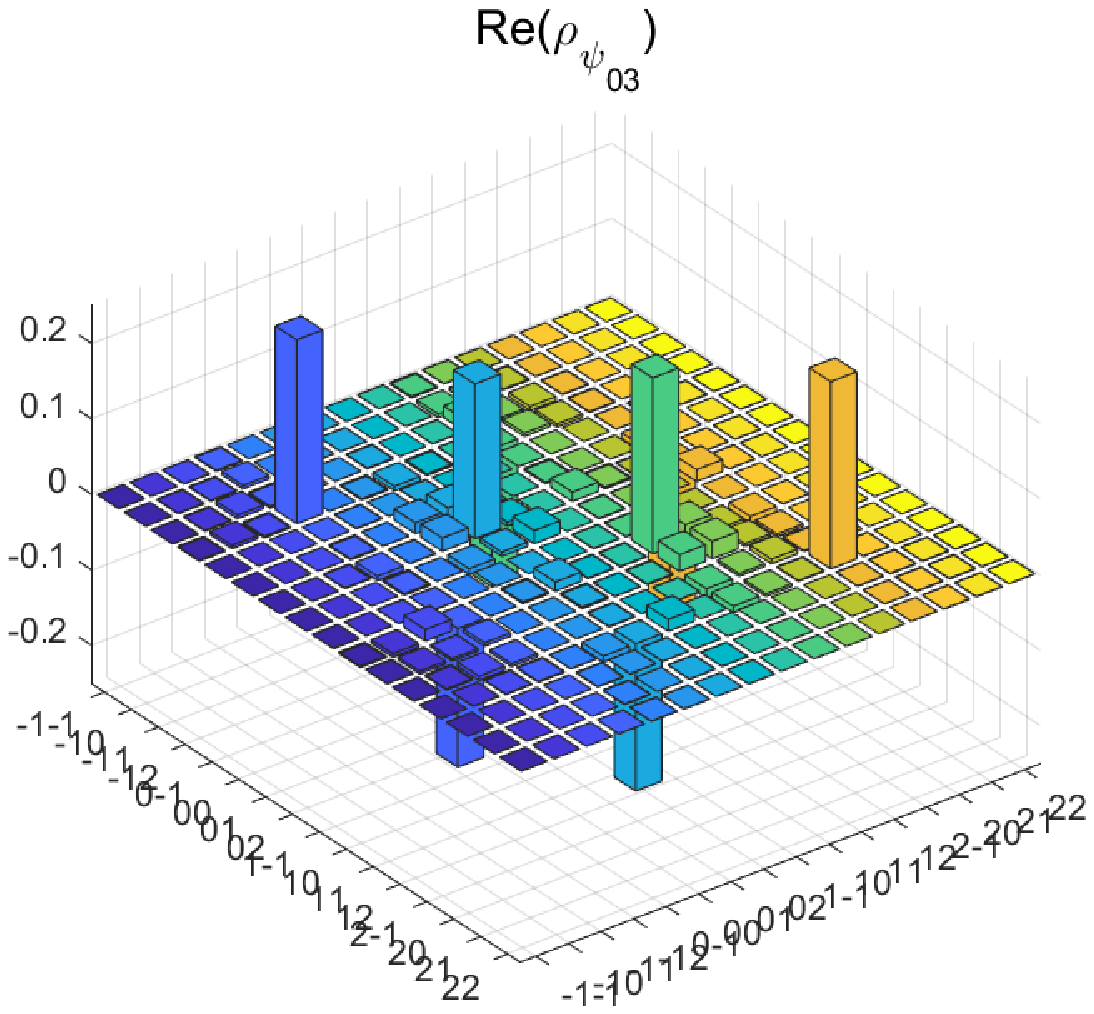}}
\subfigure[]{
\label{Fig6.sub.8}
\includegraphics[width=0.24\linewidth]{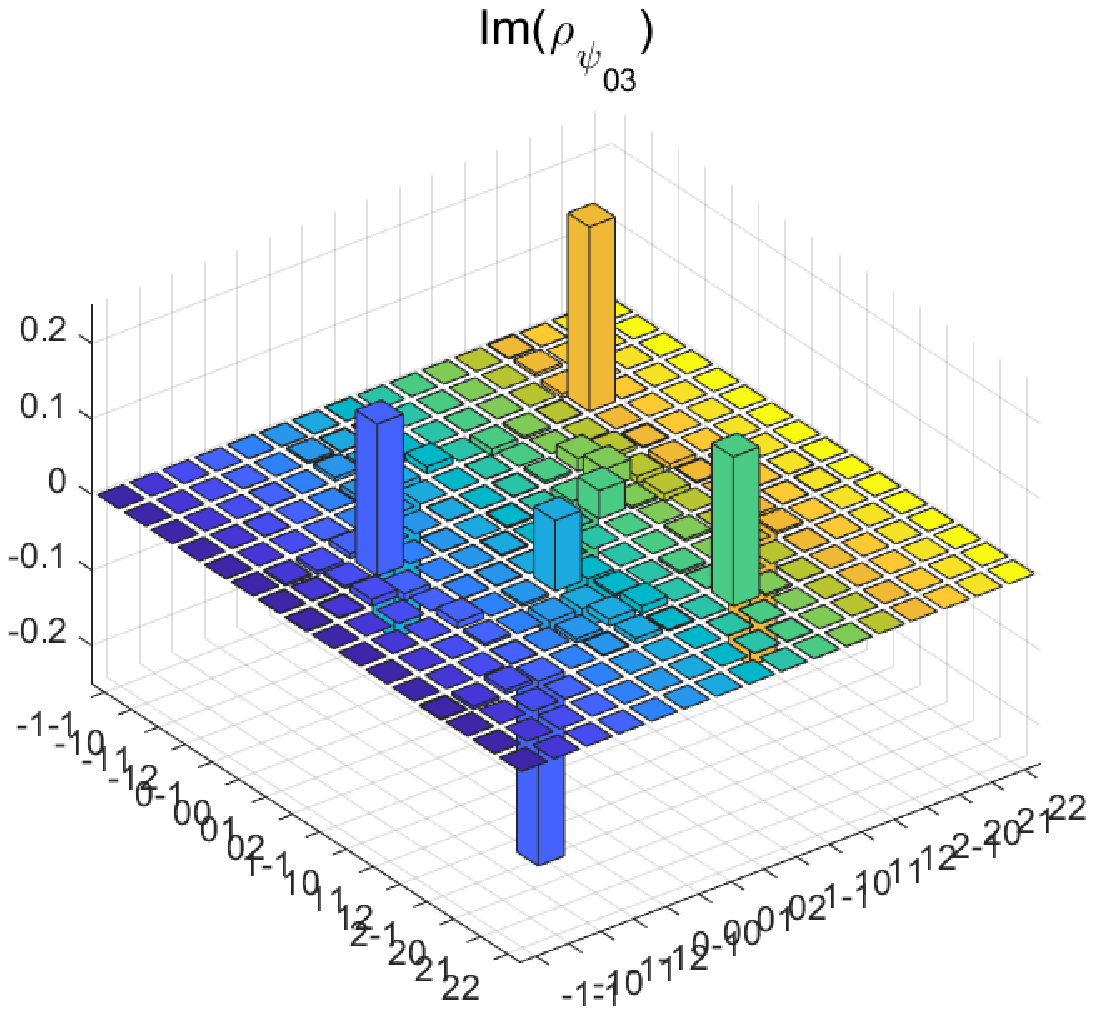}}
\subfigure[]{
\label{Fig6.sub.9}
\includegraphics[width=0.24\linewidth]{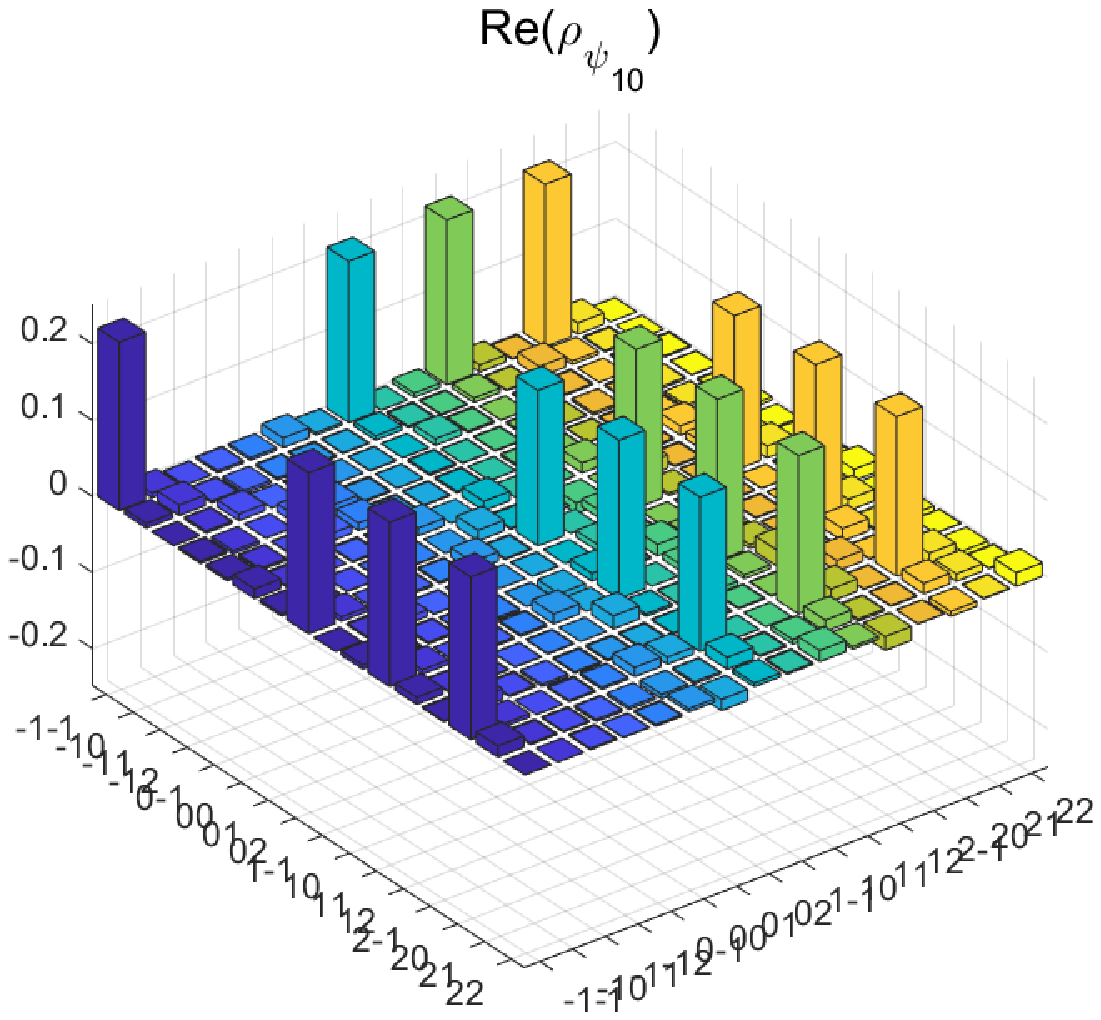}}
\subfigure[]{
\label{Fig6.sub.10}
\includegraphics[width=0.24\linewidth]{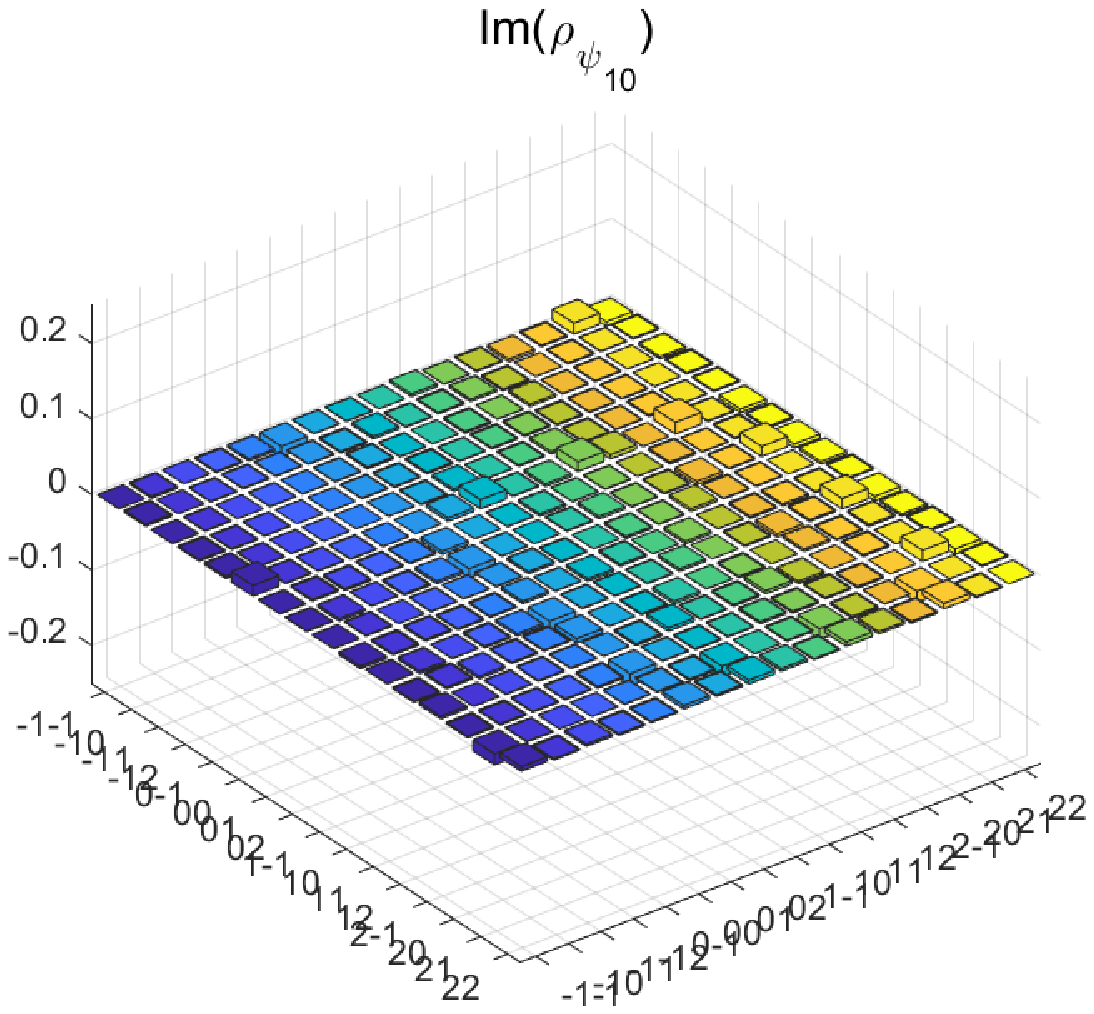}}
\subfigure[]{
\label{Fig6.sub.11}
\includegraphics[width=0.24\linewidth]{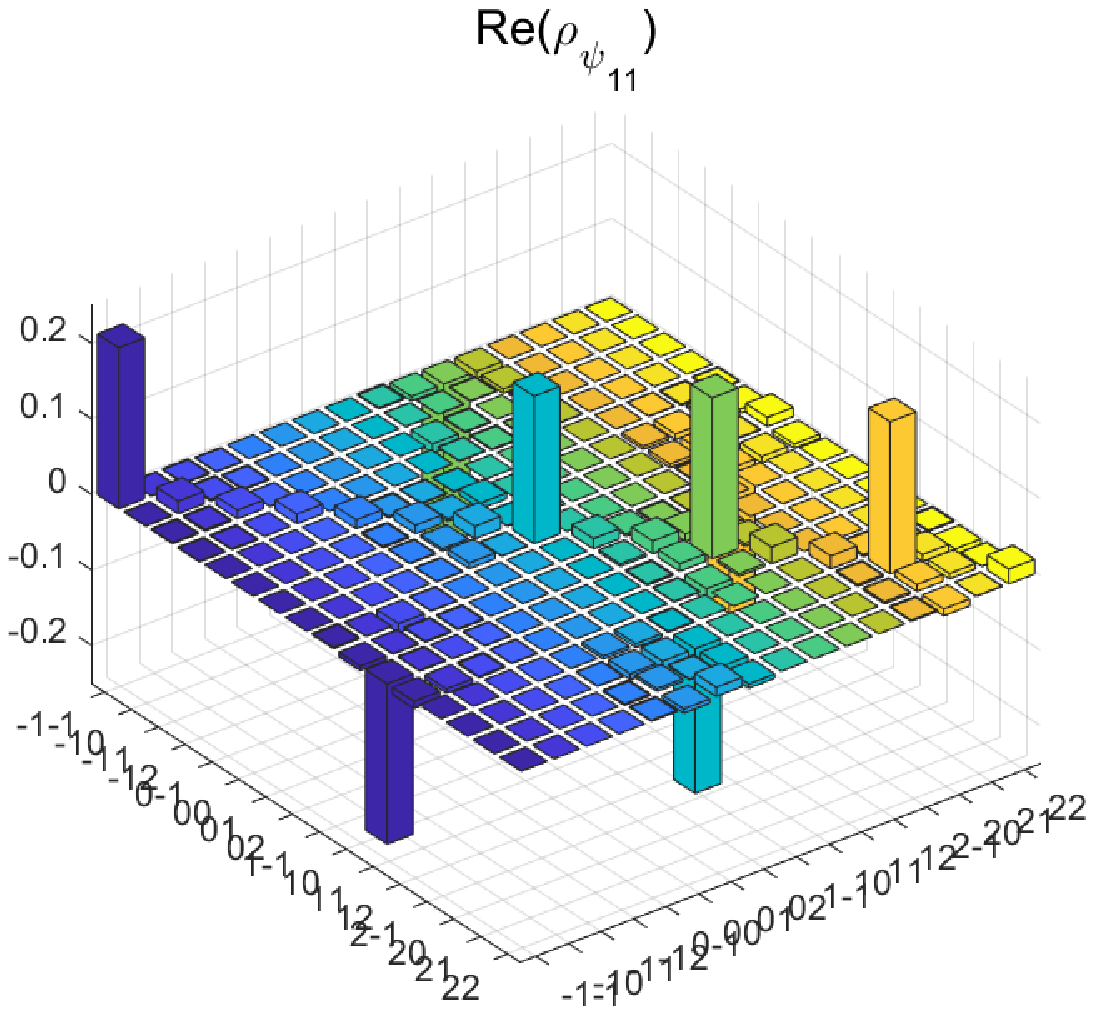}}
\subfigure[]{
\label{Fig6.sub.12}
\includegraphics[width=0.24\linewidth]{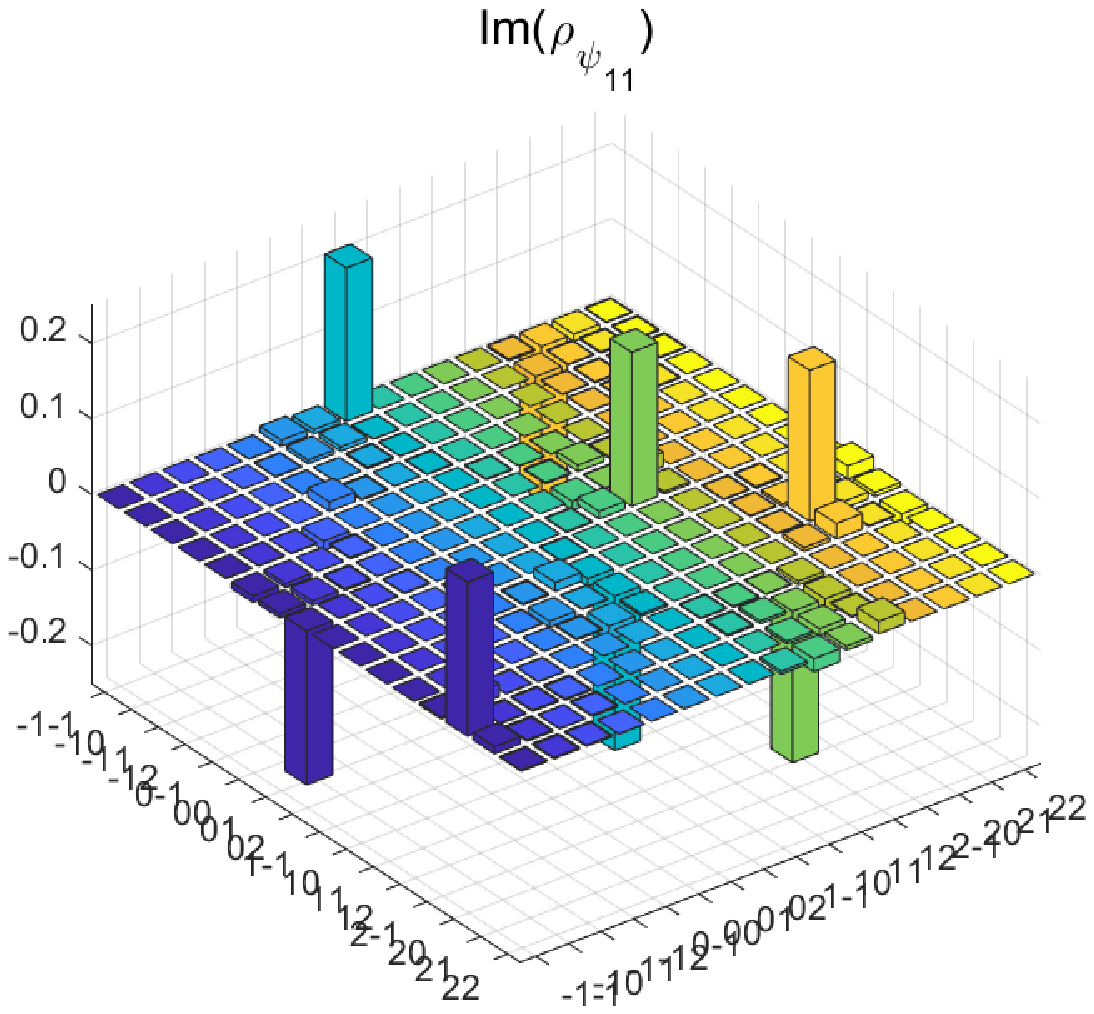}}
\subfigure[]{
\label{Fig6.sub.13}
\includegraphics[width=0.24\linewidth]{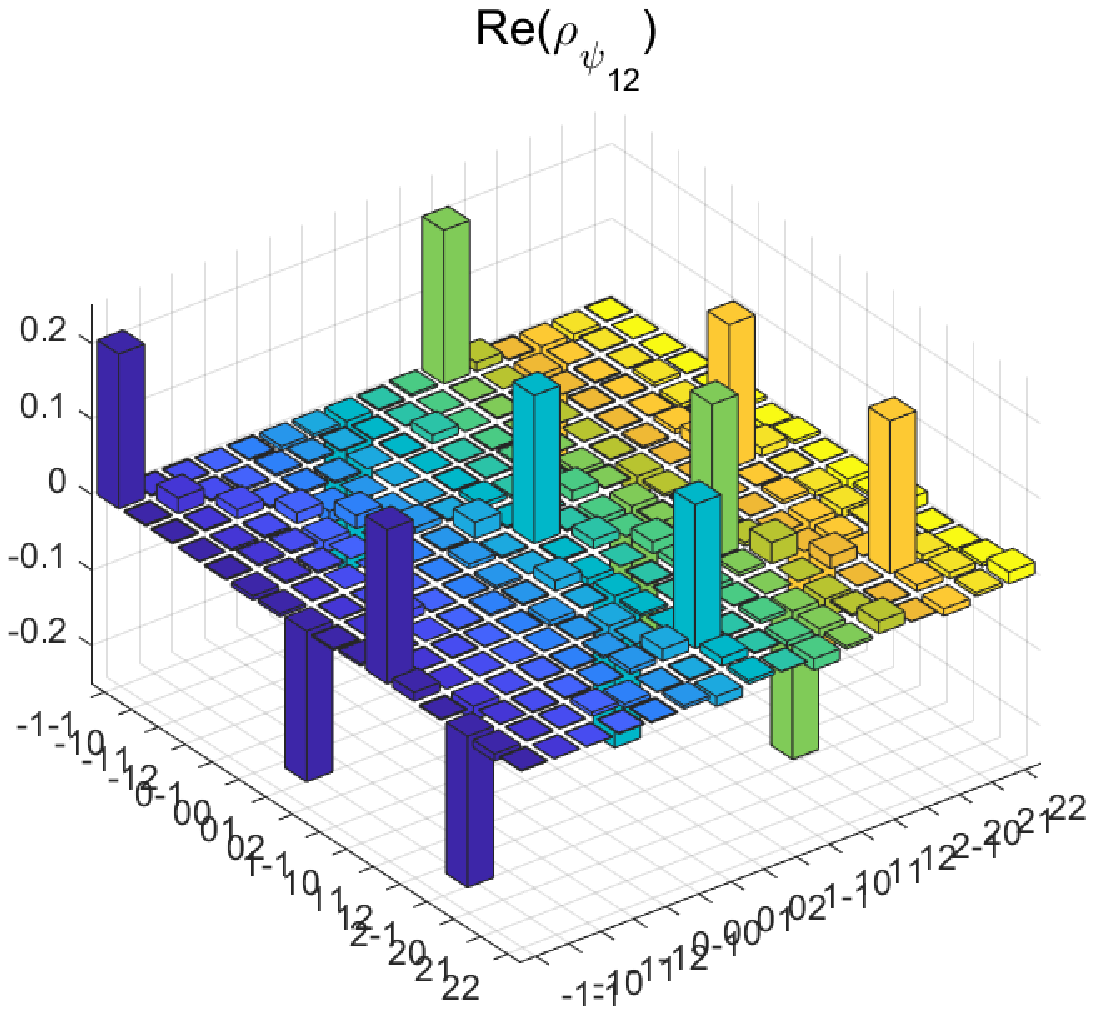}}
\subfigure[]{
\label{Fig6.sub.14}
\includegraphics[width=0.24\linewidth]{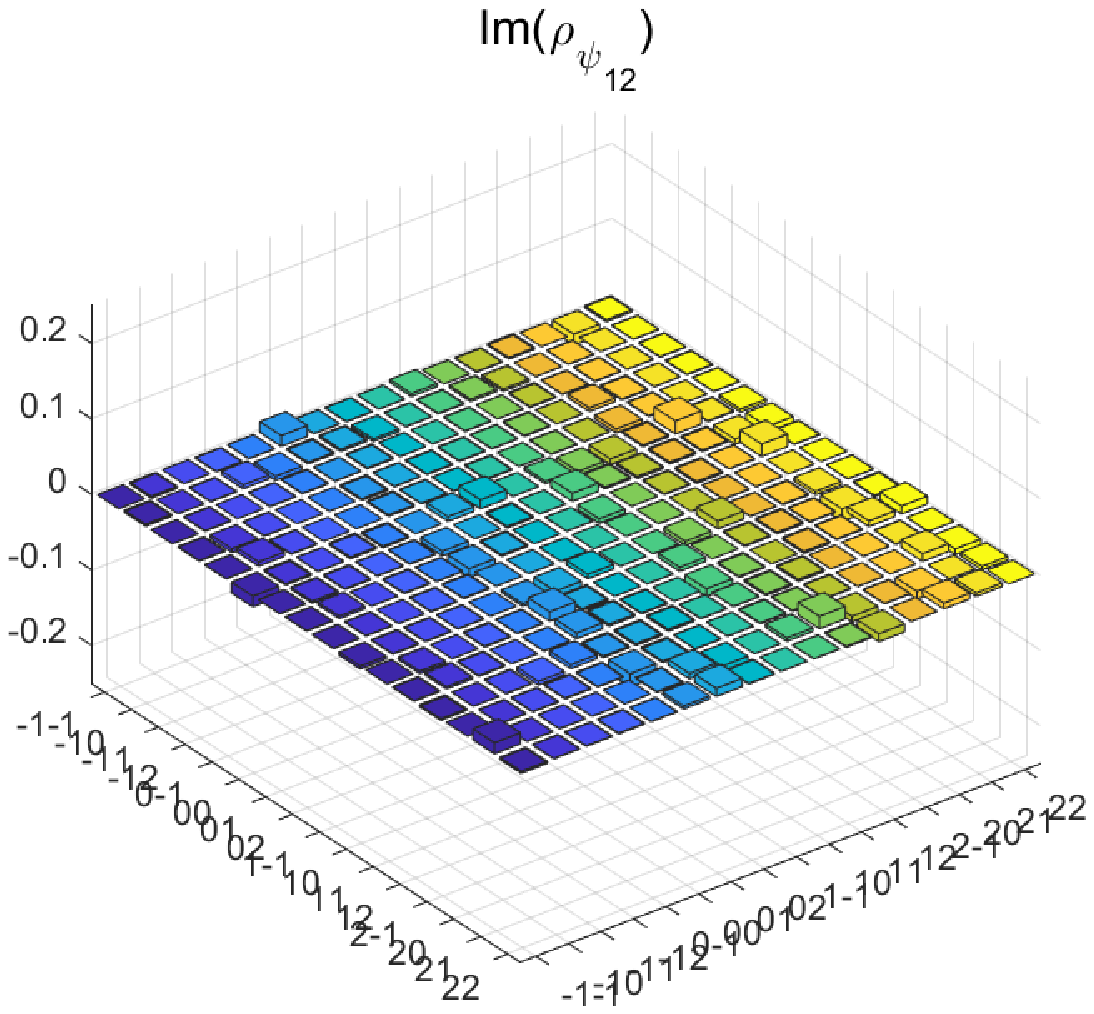}}
\subfigure[]{
\label{Fig6.sub.15}
\includegraphics[width=0.24\linewidth]{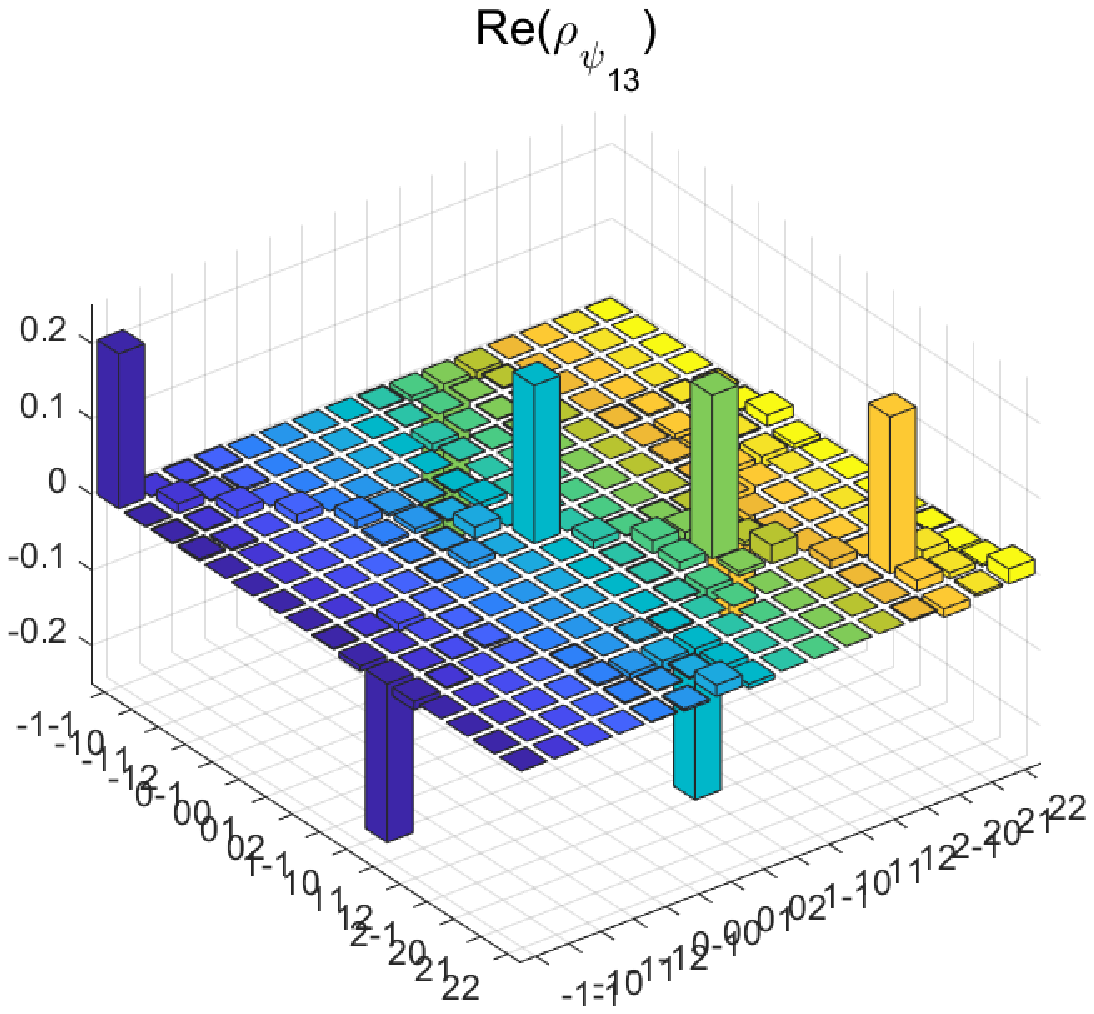}}
\subfigure[]{
\label{Fig6.sub.16}
\includegraphics[width=0.24\linewidth]{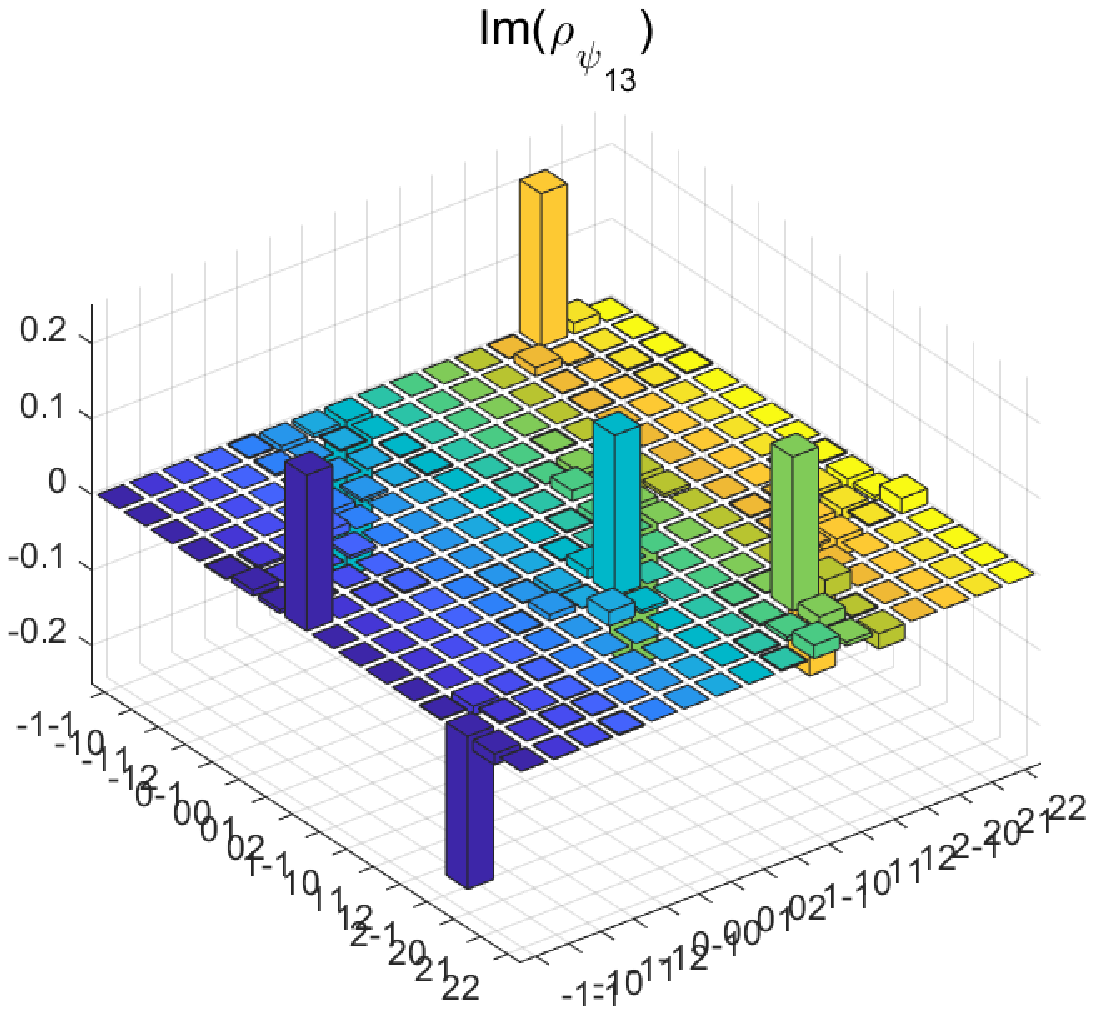}}
\end{figure*}
\addtocounter{figure}{-1}
\begin{figure*}[!t]
\addtocounter{figure}{1}
\centering
\subfigure[]{
\label{Fig6.sub.5}
\includegraphics[width=0.24\linewidth]{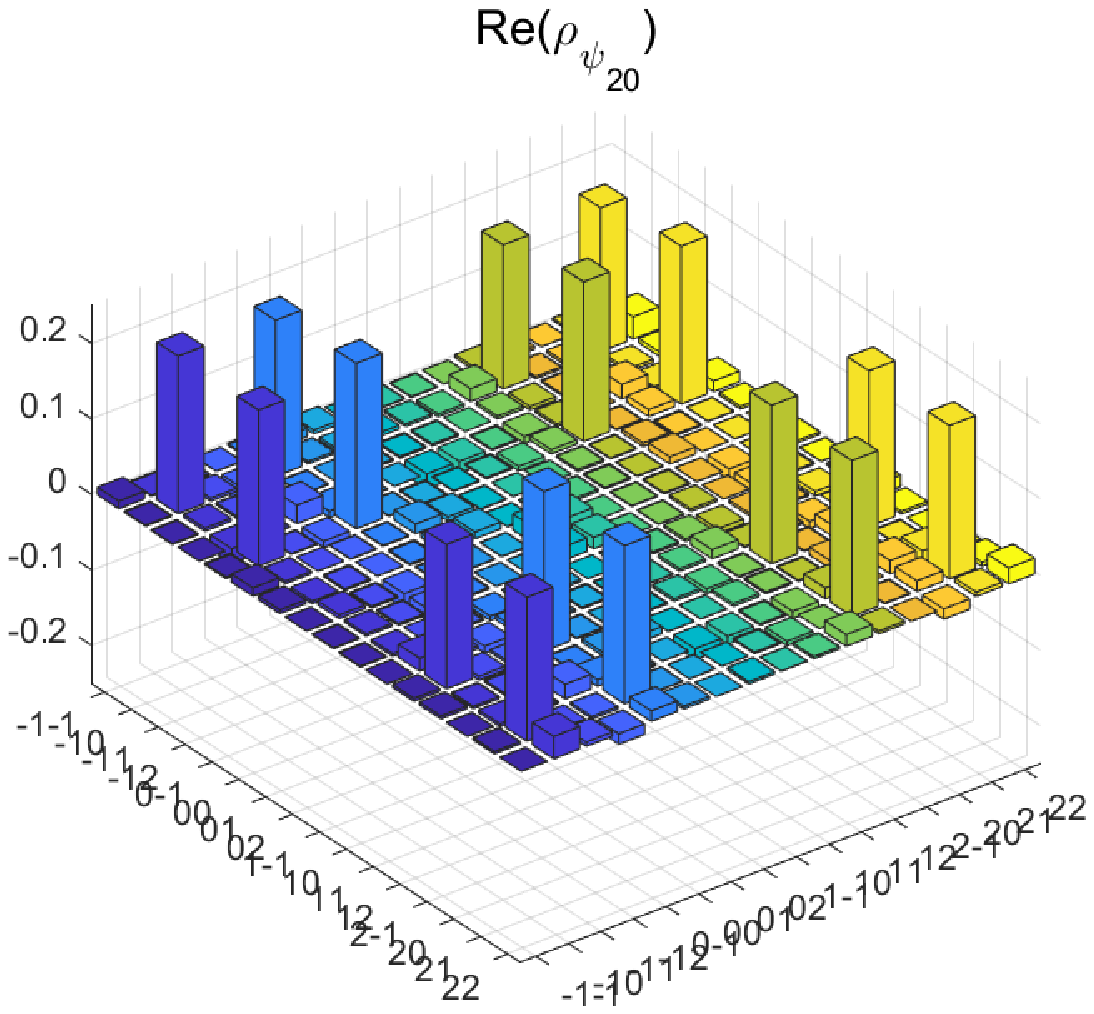}}
\subfigure[]{
\label{Fig6.sub.6}
\includegraphics[width=0.24\linewidth]{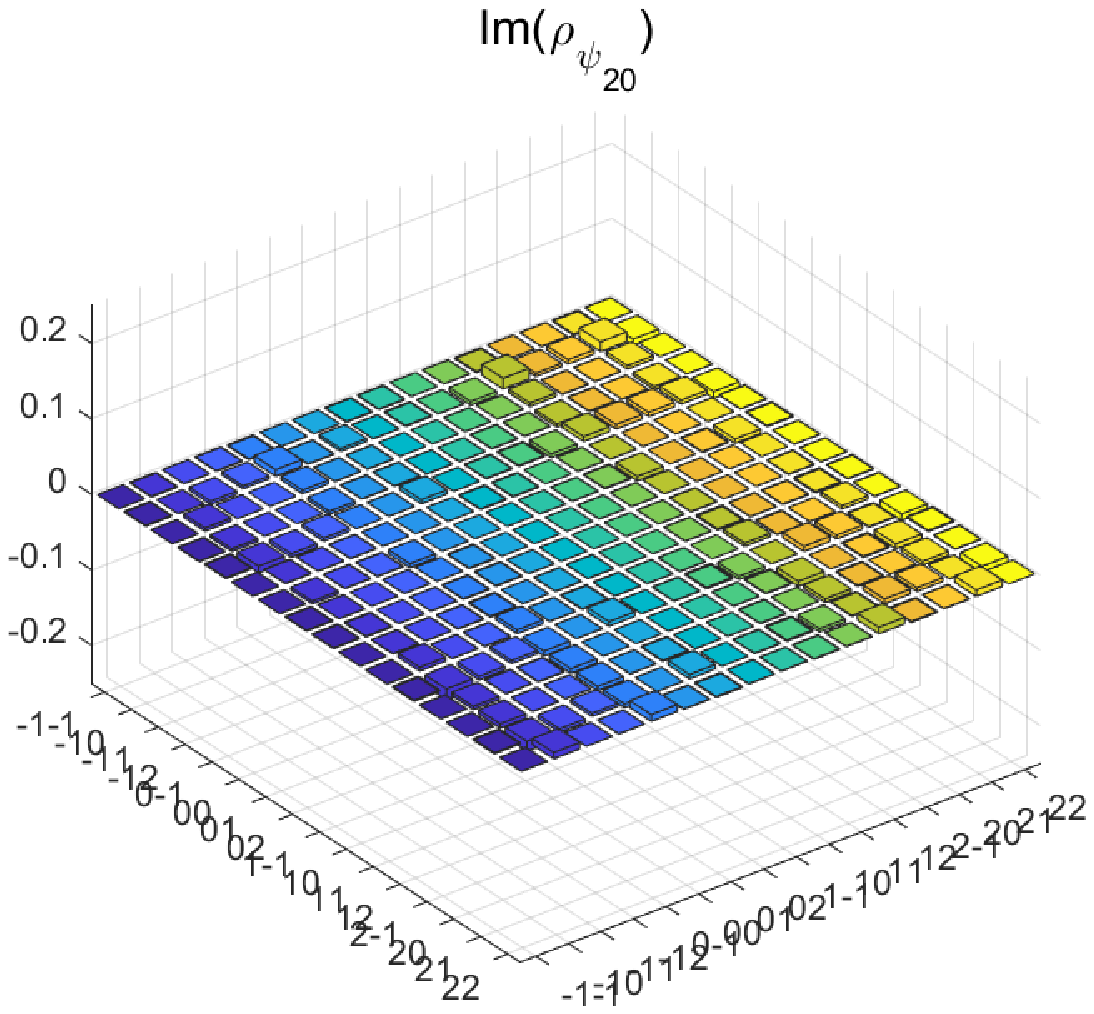}}
\subfigure[]{
\label{Fig6.sub.1}
\includegraphics[width=0.24\linewidth]{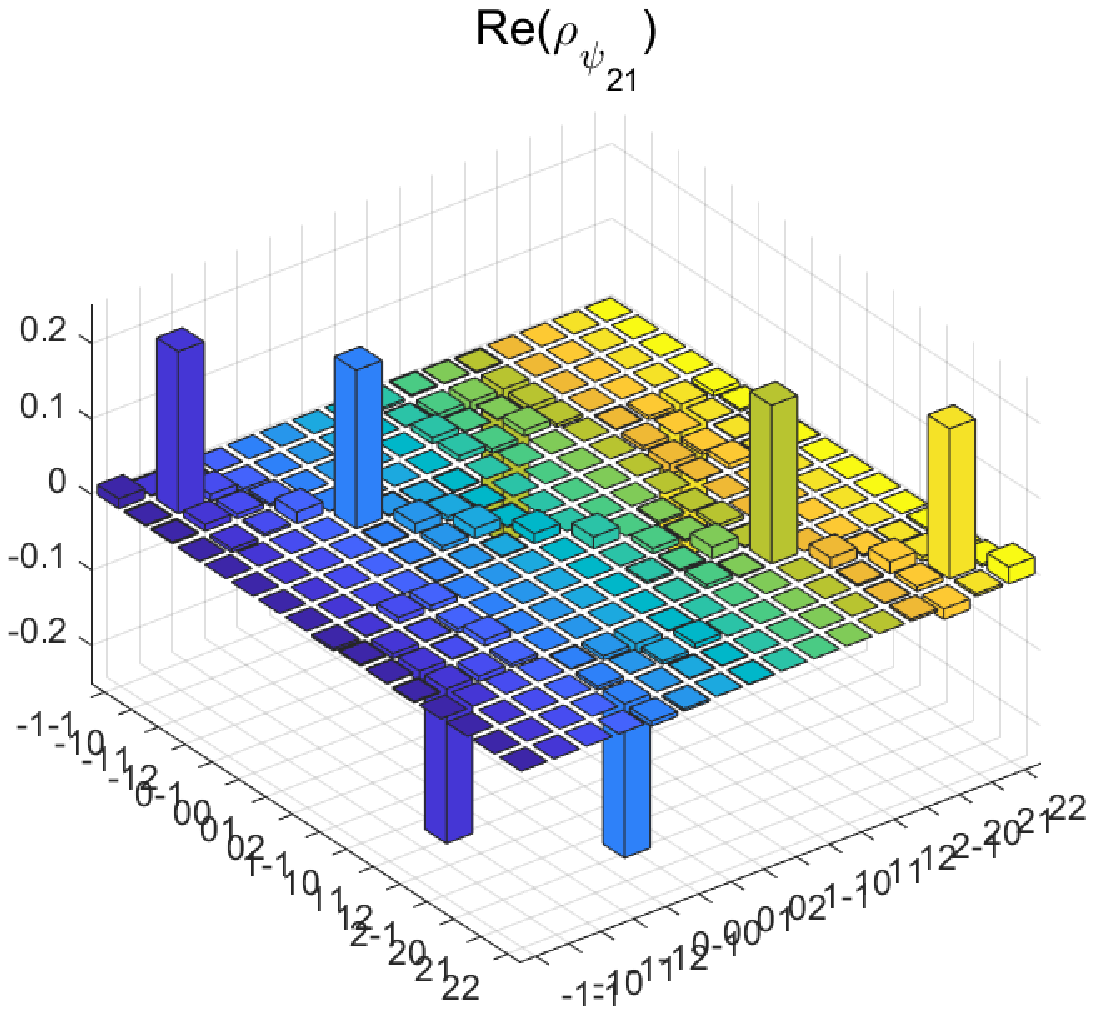}}
\subfigure[]{
\label{Fig6.sub.2}
\includegraphics[width=0.24\linewidth]{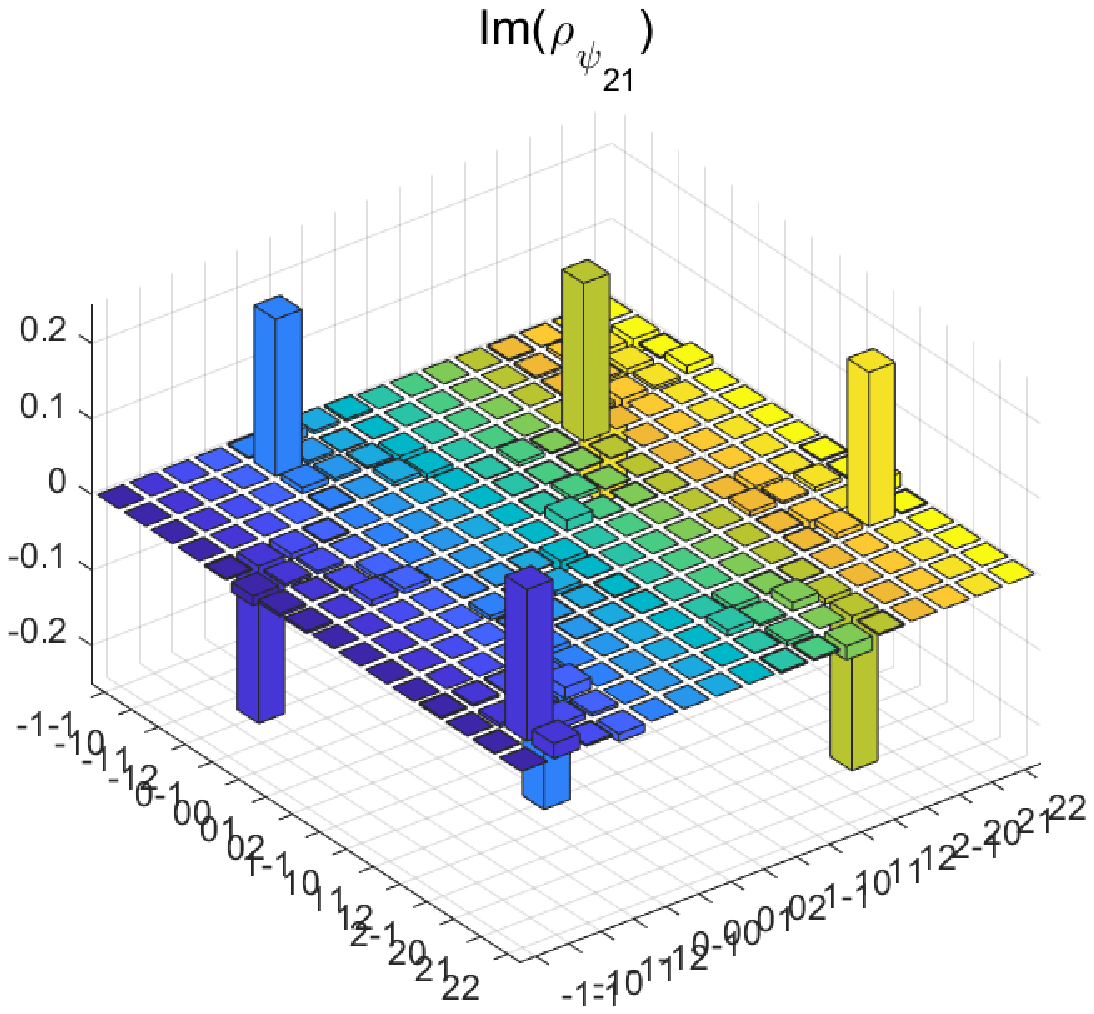}}
\subfigure[]{
\label{Fig6.sub.1}
\includegraphics[width=0.24\linewidth]{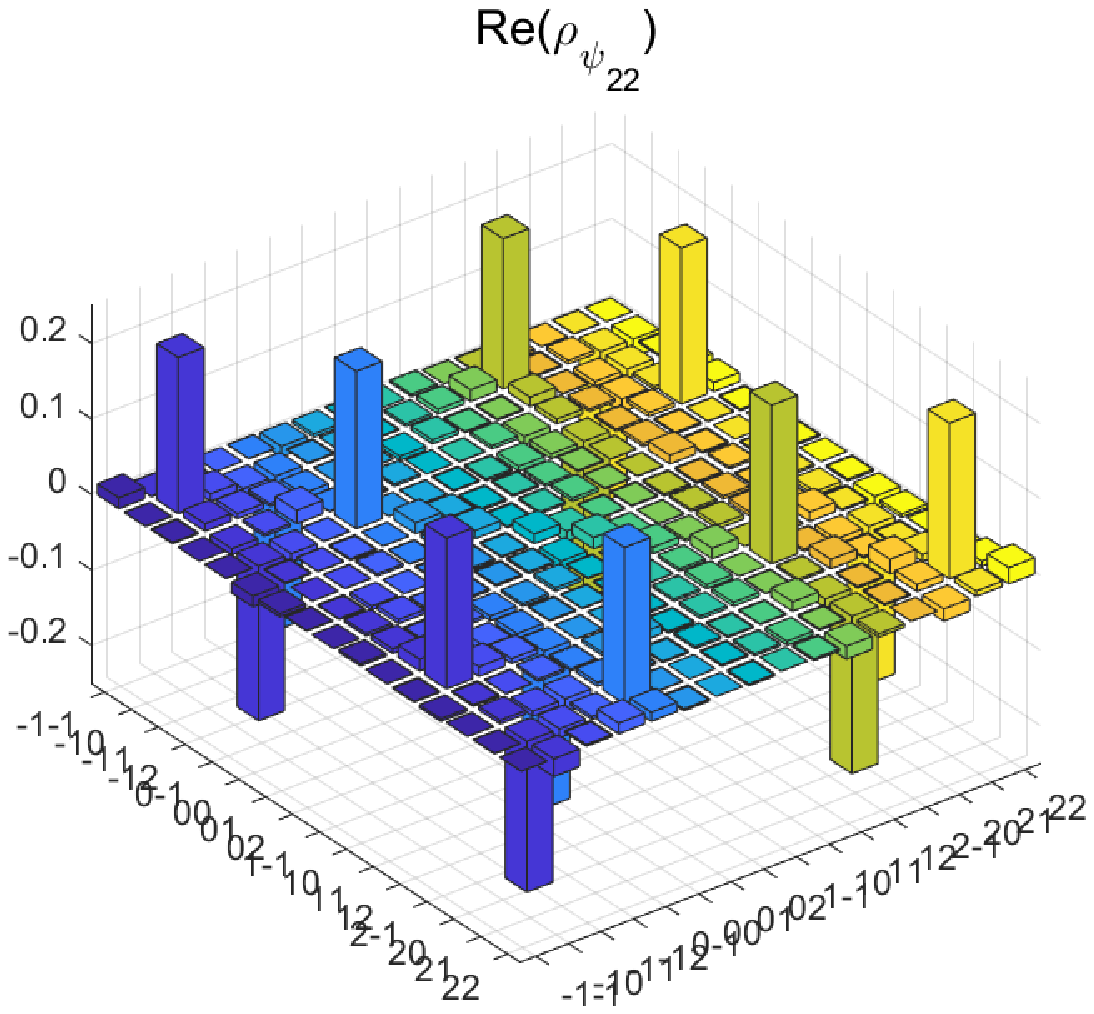}}
\subfigure[]{
\label{Fig6.sub.2}
\includegraphics[width=0.24\linewidth]{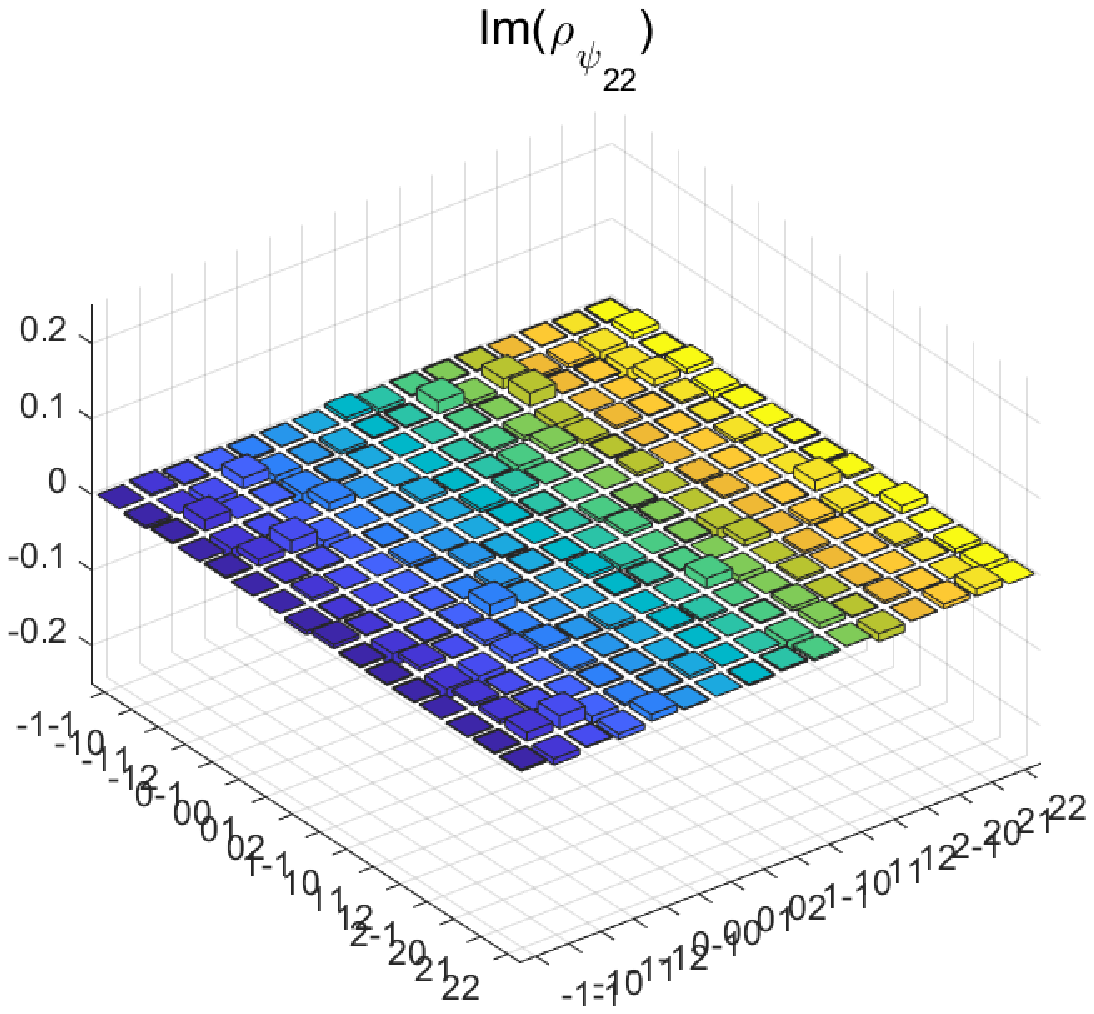}}
\subfigure[]{
\label{Fig6.sub.1}
\includegraphics[width=0.24\linewidth]{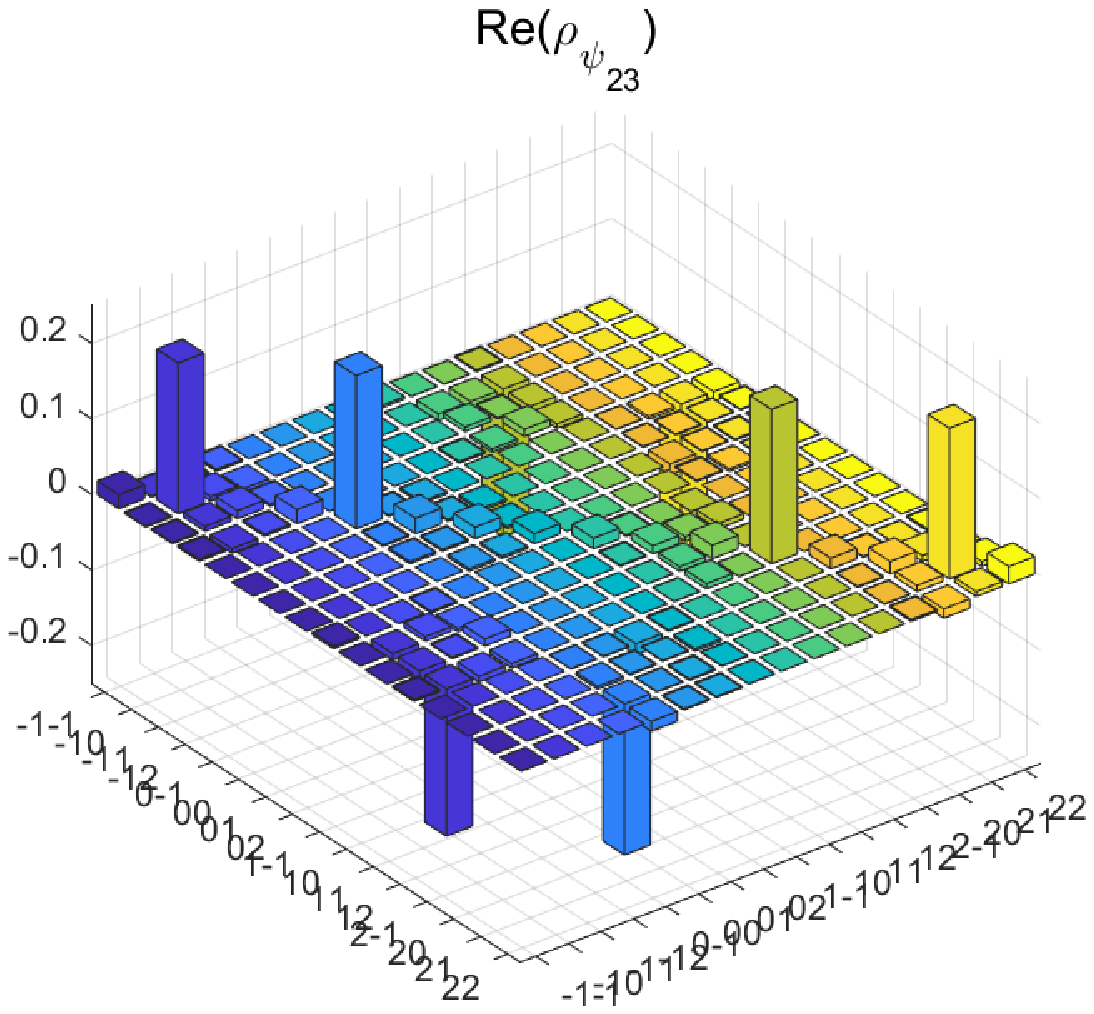}}
\subfigure[]{
\label{Fig6.sub.2}
\includegraphics[width=0.24\linewidth]{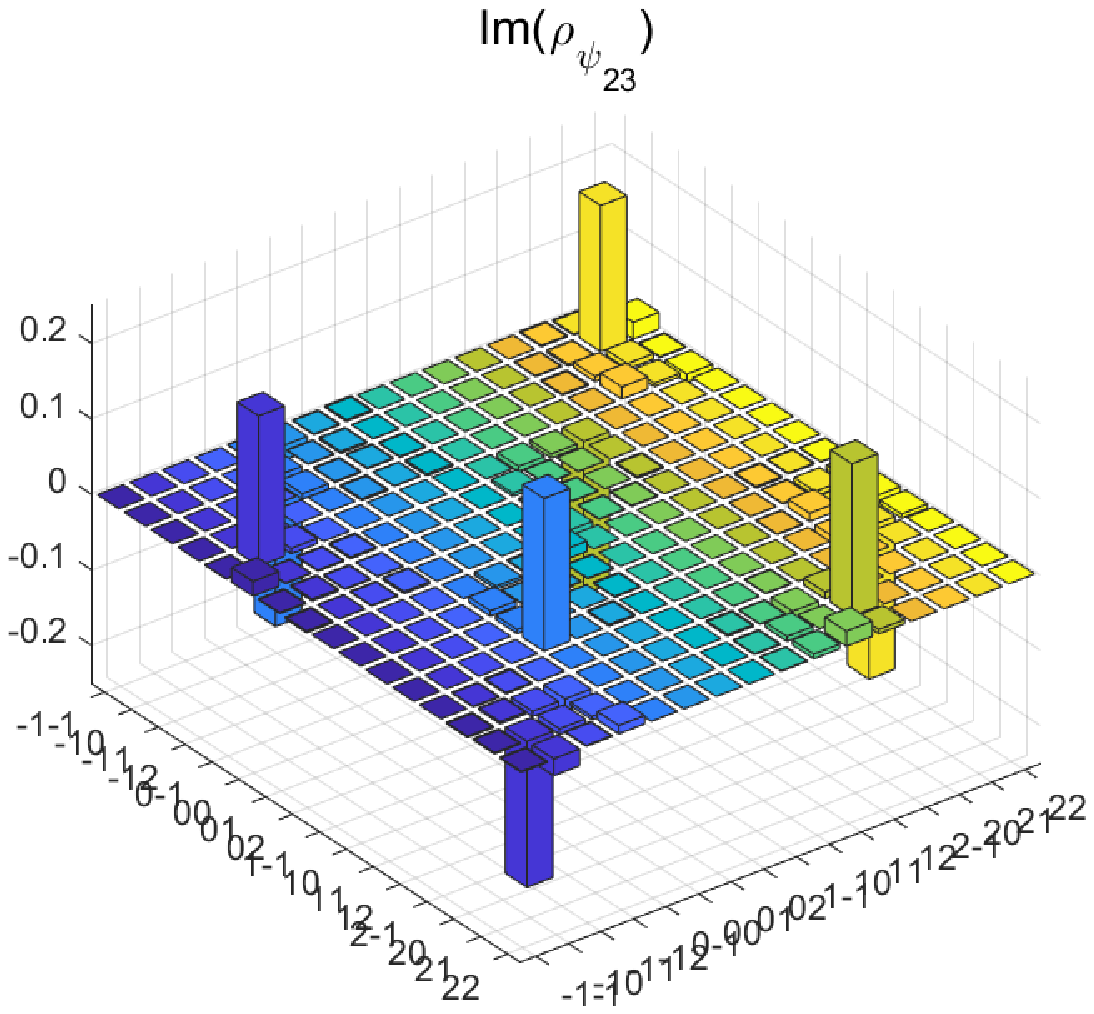}}
\subfigure[]{
\label{Fig6.sub.7}
\includegraphics[width=0.24\linewidth]{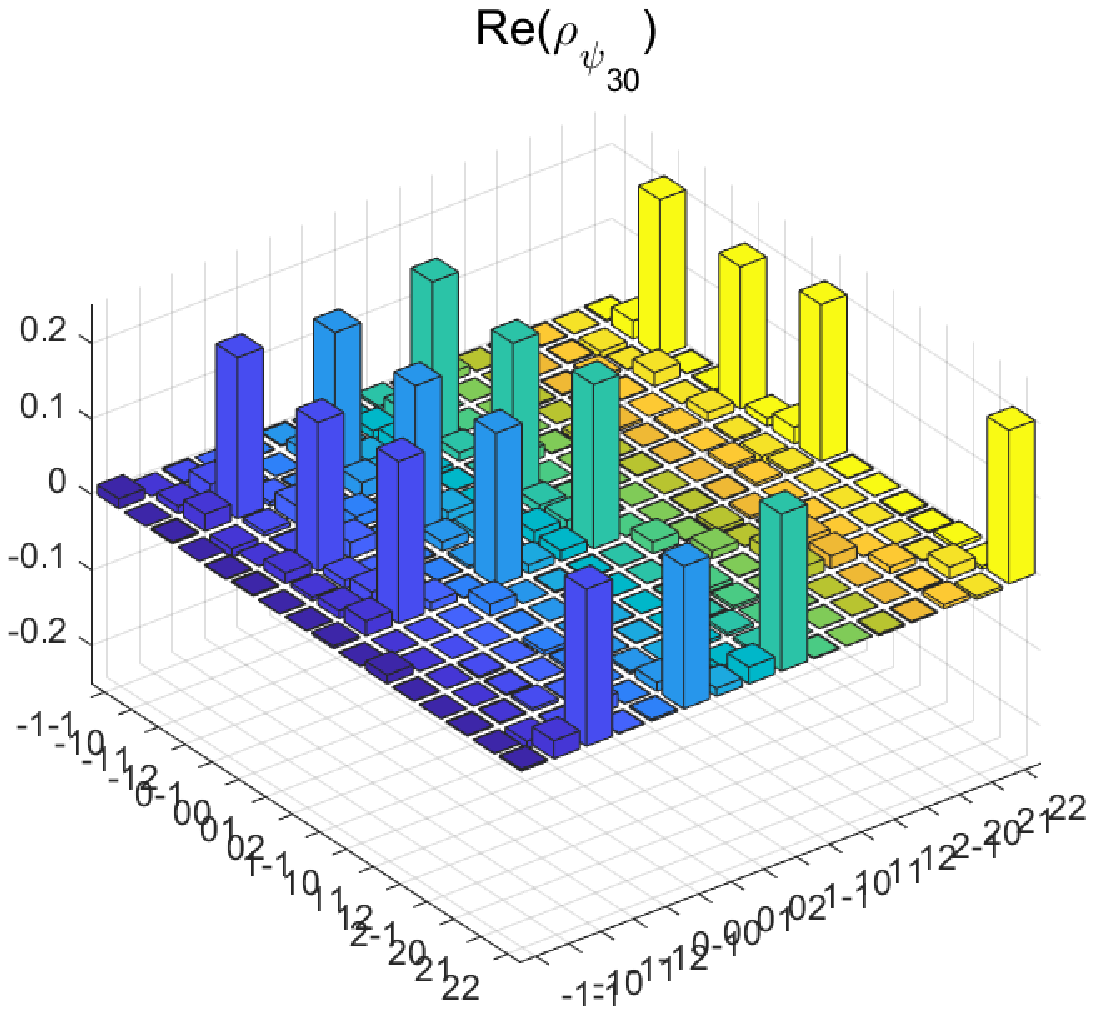}}
\subfigure[]{
\label{Fig6.sub.8}
\includegraphics[width=0.24\linewidth]{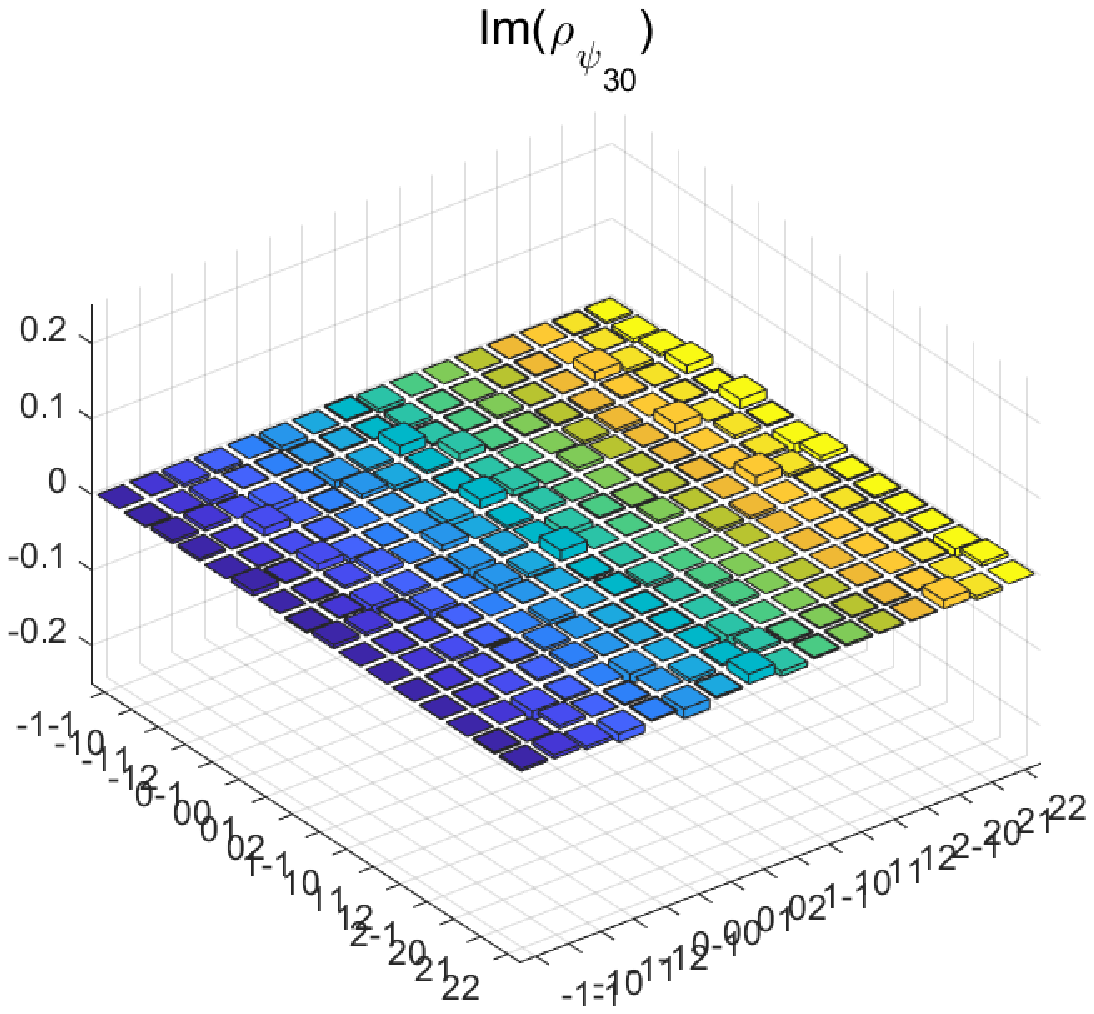}}
\subfigure[]{
\label{Fig6.sub.1}
\includegraphics[width=0.24\linewidth]{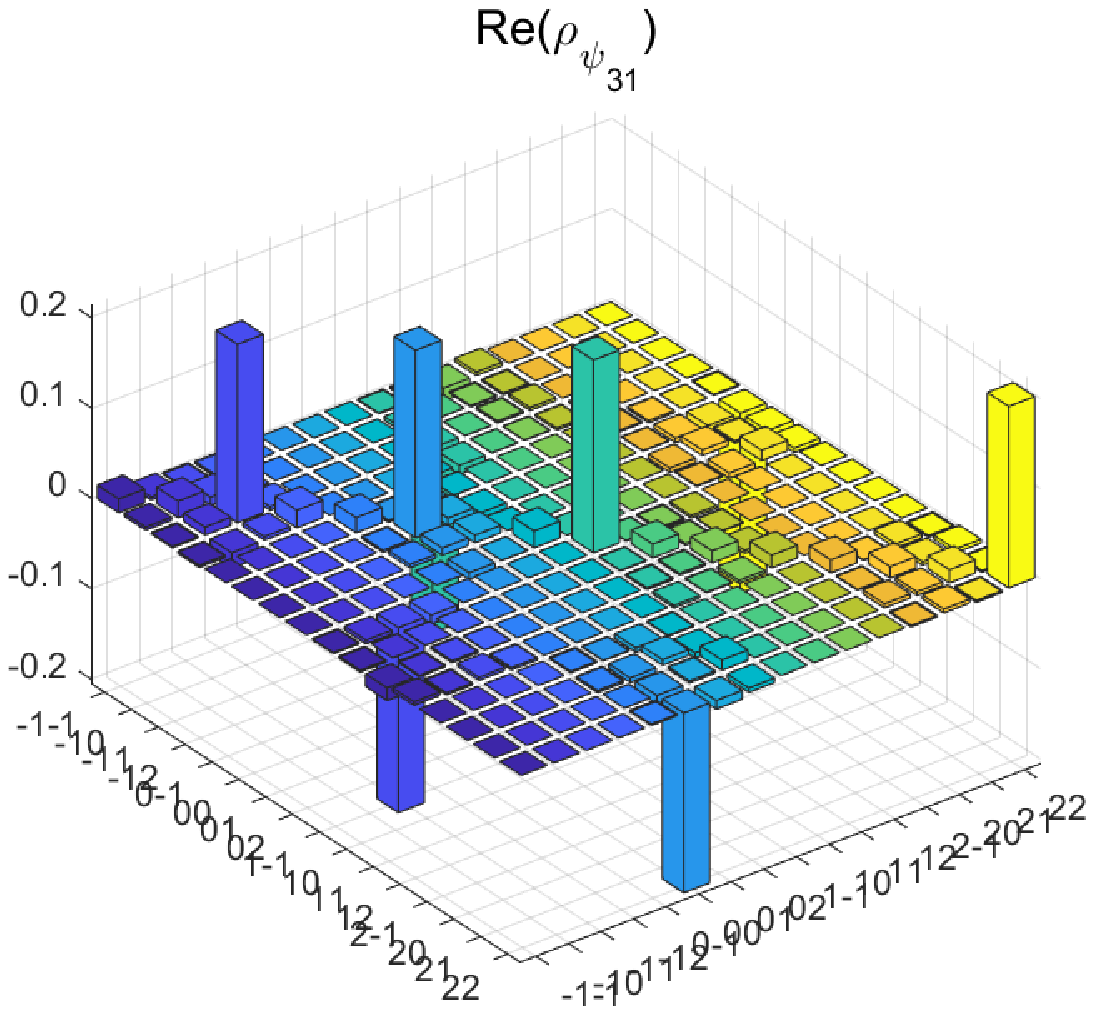}}
\subfigure[]{
\label{Fig6.sub.2}
\includegraphics[width=0.24\linewidth]{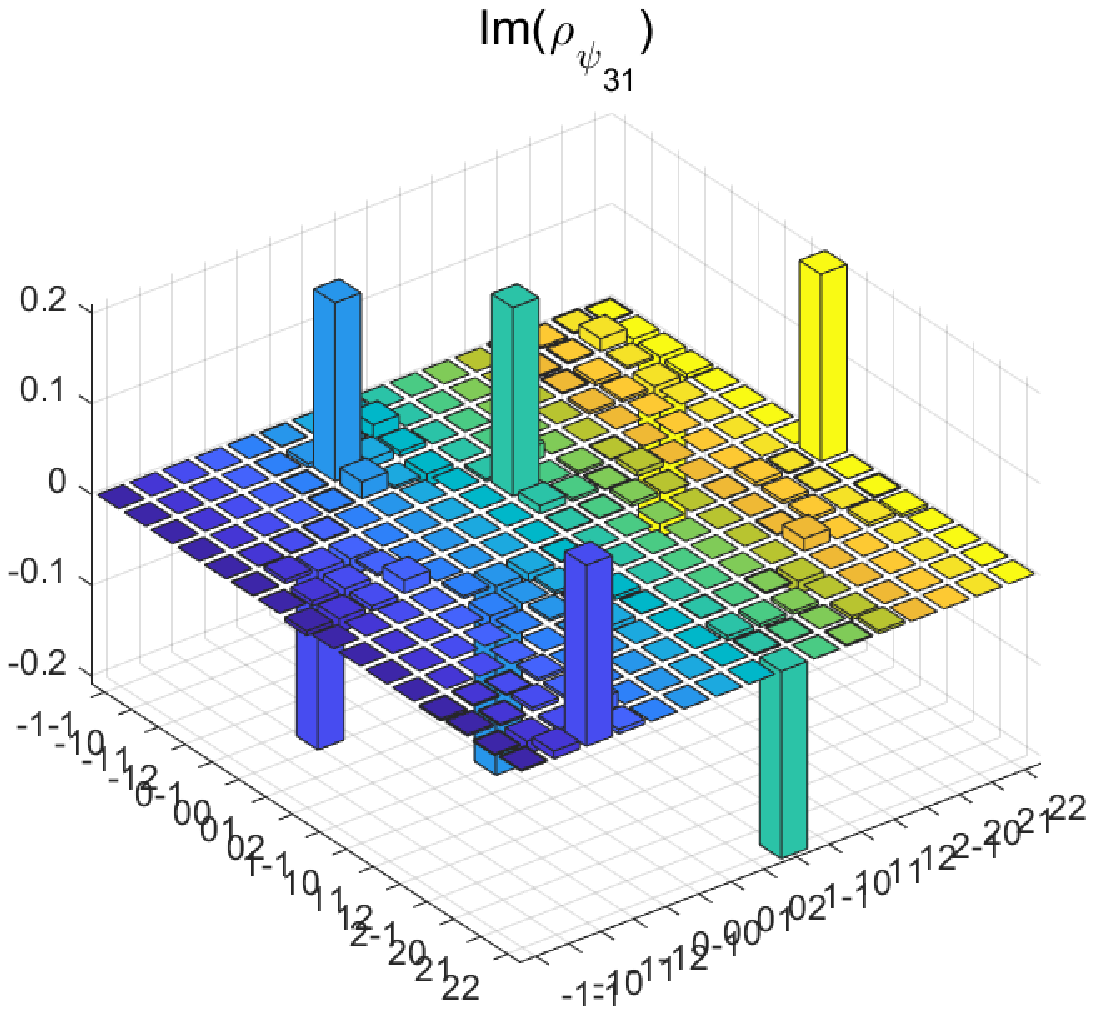}}
\subfigure[]{
\label{Fig6.sub.1}
\includegraphics[width=0.24\linewidth]{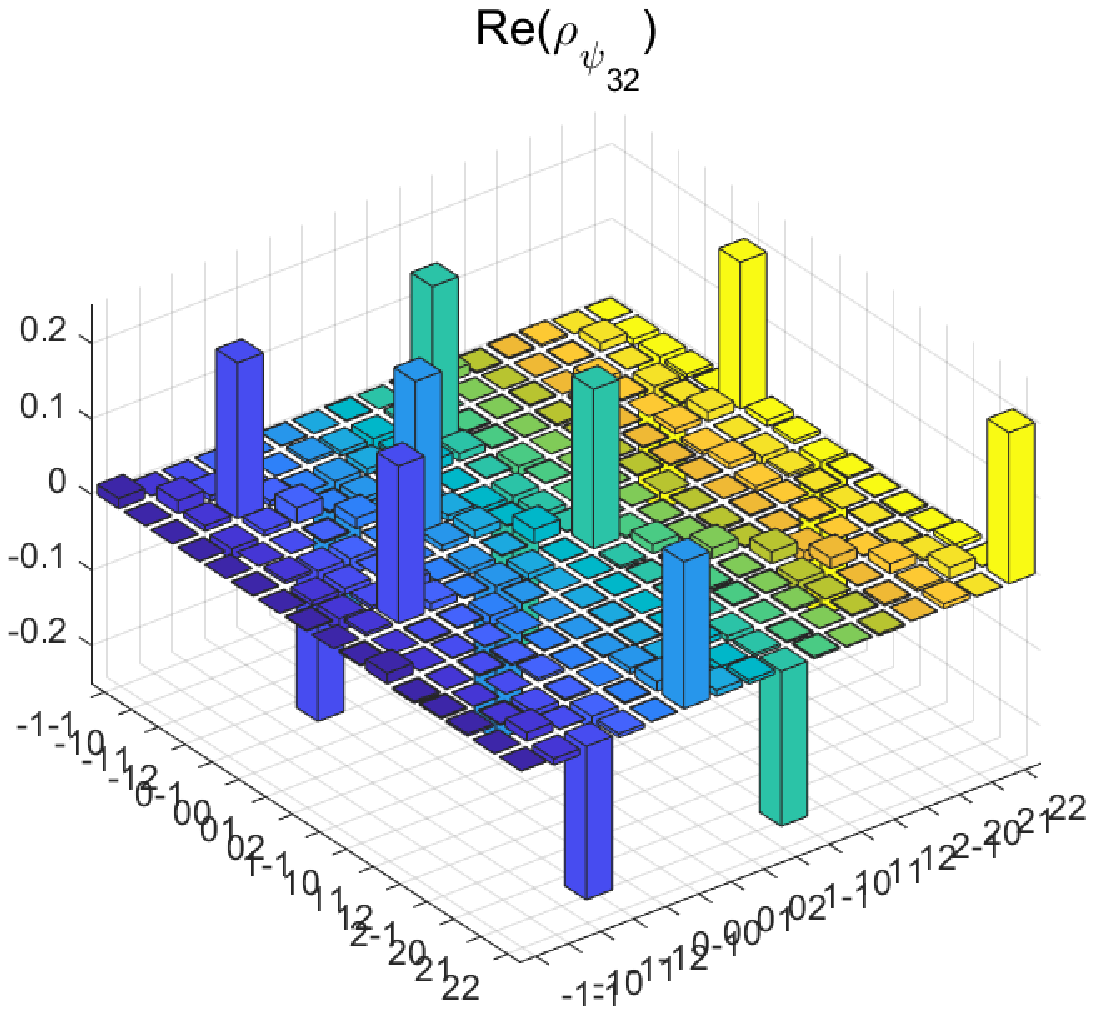}}
\subfigure[]{
\label{Fig6.sub.2}
\includegraphics[width=0.24\linewidth]{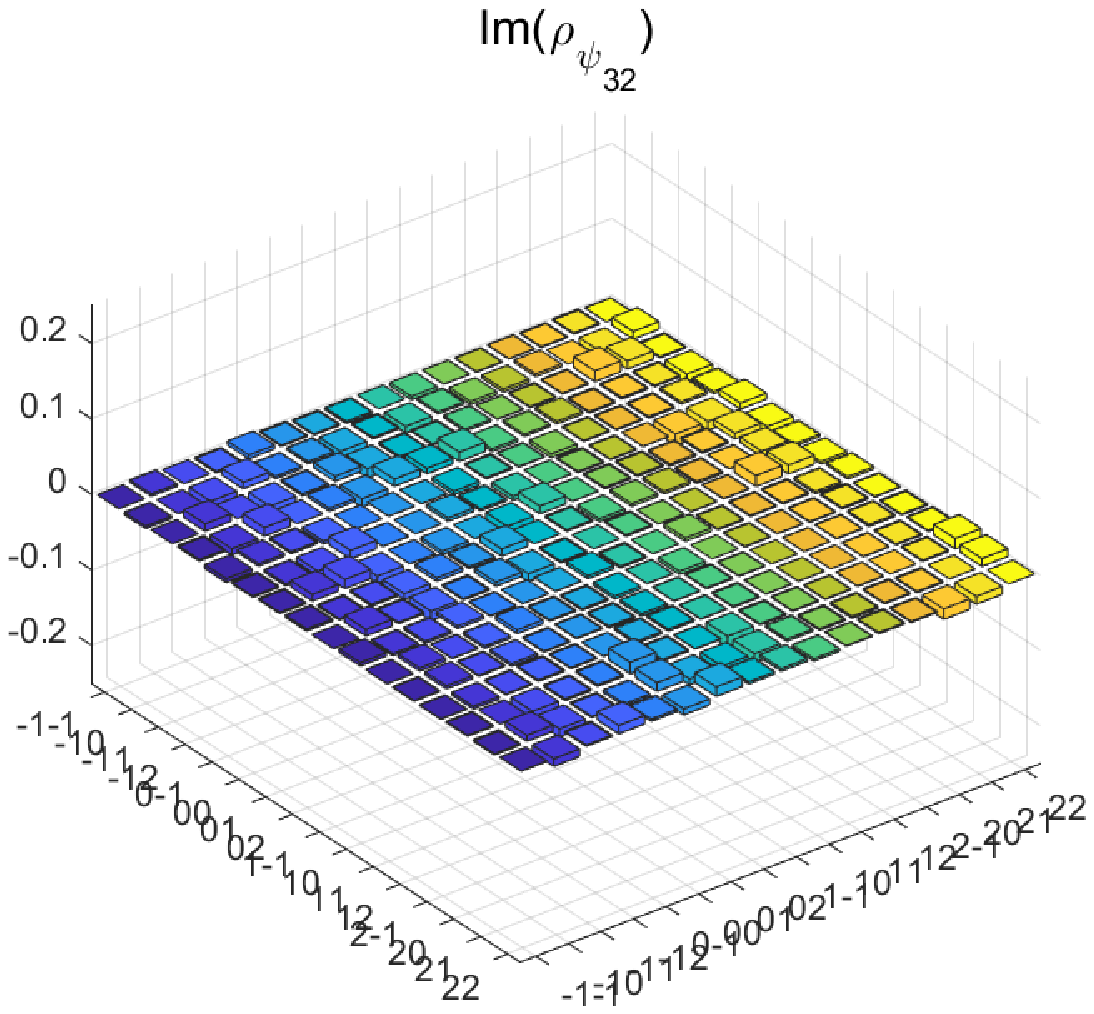}}
\subfigure[]{
\label{Fig6.sub.1}
\includegraphics[width=0.24\linewidth]{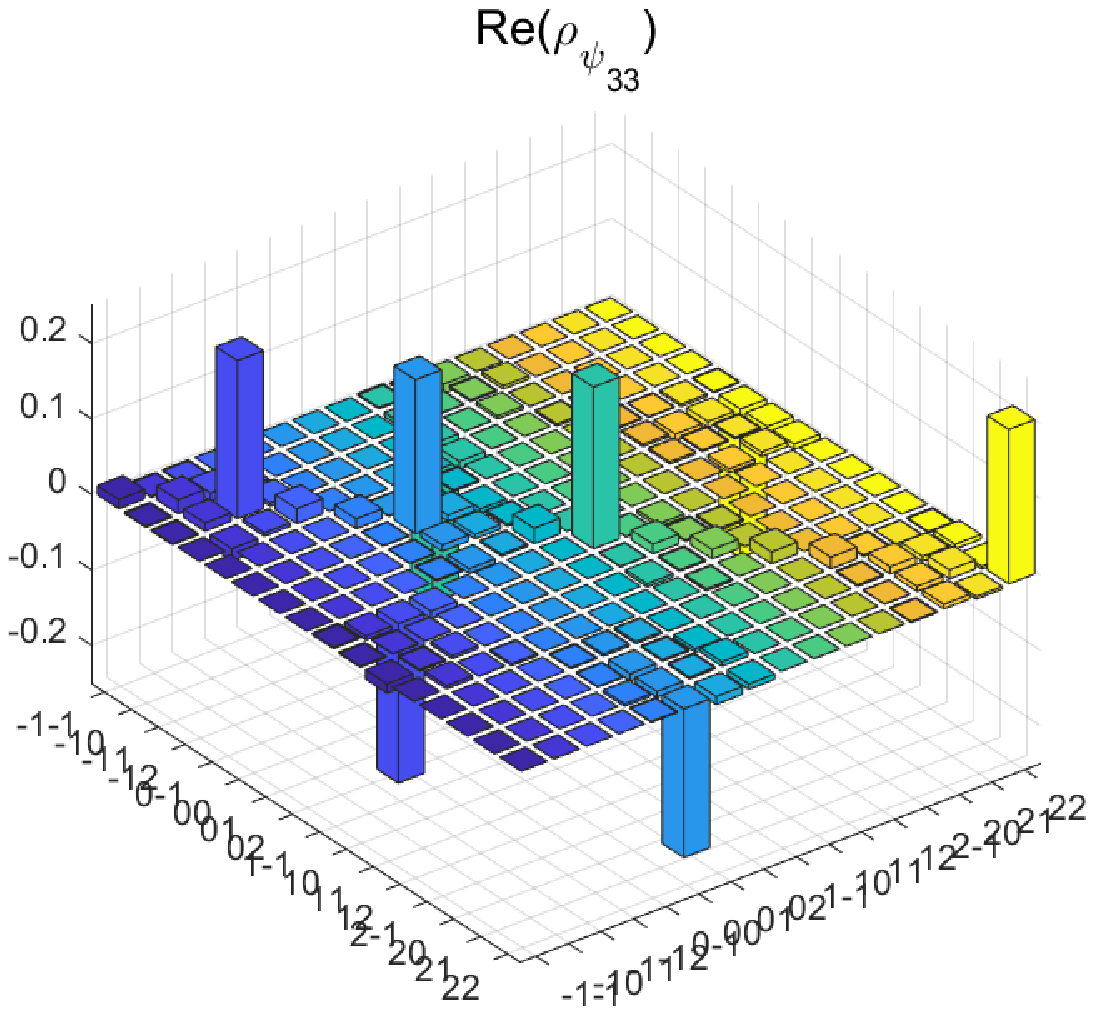}}
\subfigure[]{
\label{Fig6.sub.2}
\includegraphics[width=0.24\linewidth]{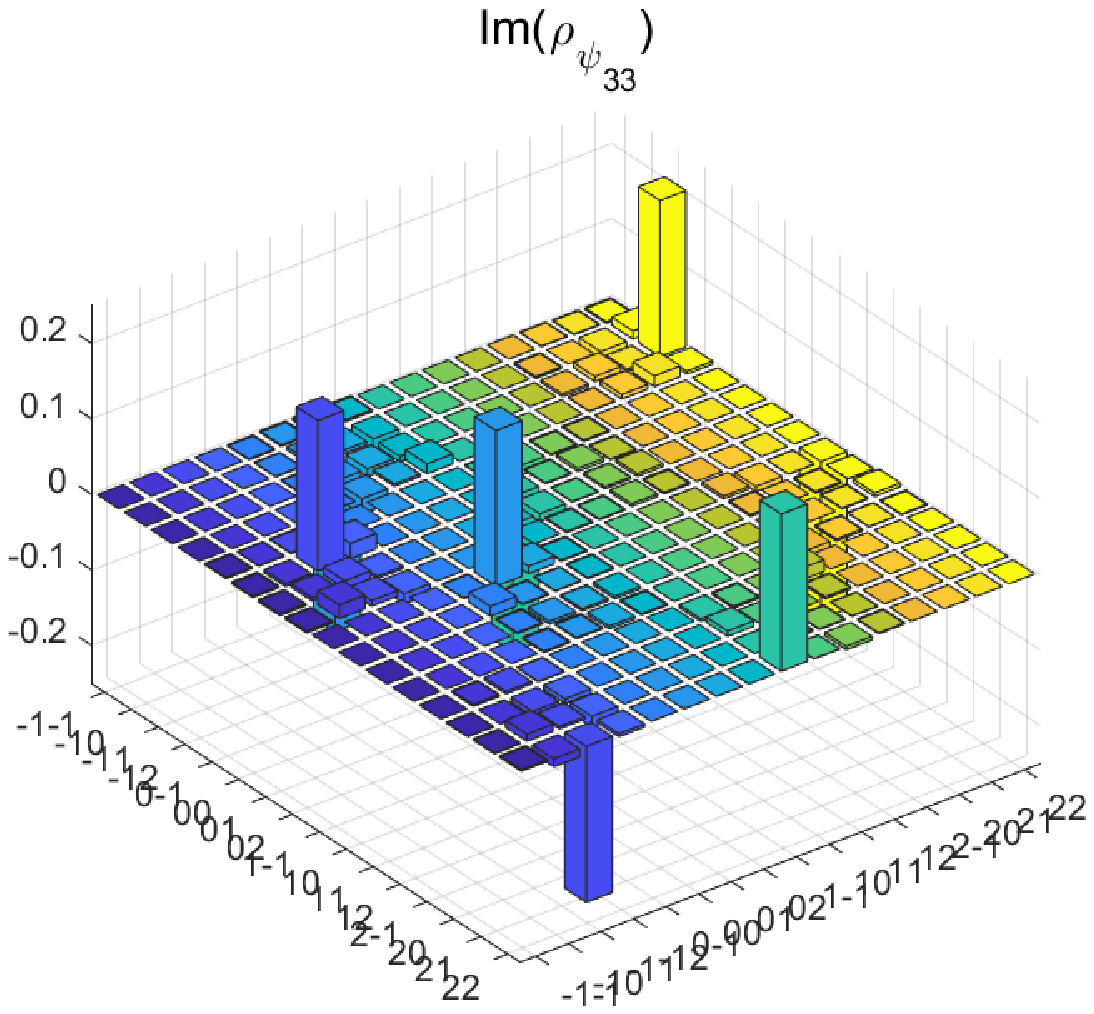}}
\caption{The estimated density matrices for 16 Bell states.}
\label{figure_6}
\end{figure*}
\end{document}